\documentstyle[11pt,emulateapj,epsfig]{article}
\slugcomment{To appear in the Astronomical Journal, 2003}

\lefthead{Cohen \etal}
\righthead{$B$-Band Parallel Survey}

\newcommand{\etal}	{{et\thinspace al.} }
\newcommand{\re}	{{$r_e$} }
\newcommand{\cf}	{{\it cf.} }
\newcommand{\ie}	{{\it i.e.}, }
\newcommand{\eg}	{{\it e.g.}, }
\newcommand{\zmed}	{{z_{med}} }
\newcommand{\magarc}	{{\ mag\ arcsec$^{-2}$} }
\newcommand{\Bj}	{{b_J} }

\begin{document}

\title{The {\it Hubble Space Telescope\altaffilmark{1} WFPC2} $B$--Band
Parallel Survey:\\ A Study of Galaxy Morphology for $18\leq$$B$$\leq27$ mag}

\author{Seth H. Cohen, Rogier A. Windhorst, Stephen C. Odewahn and Claudia A.
Chiarenza}
\affil{Department of Physics and Astronomy, Arizona State University, 
PO Box 871504, Tempe, AZ 85287-1504}

\and
\author{Simon P. Driver}
\affil{Research School of Astronomy and Astrophysics, Australian National
University, Cotter Road, Weston Creek, ACT2600, Australia}


\altaffiltext{1}{Based on observations with the NASA/ESA {\it Hubble Space
Telescope} obtained at the Space Telescope Science Institute, which is operated
by AURA, Inc., under NASA contract NAS 5--26555.}


\begin{abstract}

We present the results of the Hubble Space Telescope $B$--Band Parallel 
Survey (BBPS). It covers 0.0370 square degrees and consists of 31 shallow
(4--6 orbit), randomly selected high latitude HST WFPC2 parallel fields with
images taken in both the $B$ (F450W) and $I$ (F814W) filters. The goal of this
survey is to morphologically classify the galaxies in a homogeneous manner
and study galaxy properties as a function of type and $B$--band magnitude for
$18\lesssim b_J \lesssim23.5$ mag. The full sample contains 1800 galaxies,
370 of which are brighter than the formal statistical completeness limit of
$b_J \lesssim23.5$ mag. The galaxies are selected from the $B$--band
images and classified using an artificial neural network (ANN) galaxy classifier
on the higher S/N $I$--band images. These provide (more) reliable types for
$I \lesssim 24$ mag (or $\Bj \lesssim 26$ mag), since these $I$--band
classifications are less subject to the uncertain redshifted rest-frame
UV morphology. The ANN classification depends on the shape of the surface
brightness profile, but not on color. These results are combined with similar
(deeper) studies in the Hubble Deep Field and the deep WFPC2 field
surrounding the radio galaxy 53W002, for which galaxies have been classified
to $b_J\lesssim27$ mag. 

The galaxy counts for the combined $B$--band selected samples show adequate
statistics for a range $19\lesssim b_J \lesssim27$ mag, and are in good
agreement with other studies in the flux range where they overlap, while
showing improved statistics at the bright end. The galaxies are subdivided
into 3 morphological classes: early-types (E/S0), mid-types (Sabc) and
late-types (Sd/Irr), and the $B$--band counts are presented for each class,
as well as the total counts. The faint end of the counts is dominated by the
irregular galaxies, which have a steep count slope of
$d$log$N$$\slash$$dm\approx0.4$. These type dependent counts are compared
to models based on local luminosity functions which include the effects
of the cosmological constant, $\Omega_{\Lambda}$. The whole BBPS 
sample, along with the two deeper fields, is used to delineate the general 
trends of effective radius and ($B$--$I$) color as function of both 
morphological type and apparent magnitude for $18\lesssim b_J \lesssim27$ mag.
These properties are discussed in the context of recent redshift surveys. A 
possible explanation for the combined results is given in terms of the 
effects of $\Omega_{\Lambda}$ on the evolution of the merger rate in a 
hierarchical scenario. 

\end{abstract}




%

\section{Introduction} \label{intro}

Over the last 20 years, the galaxy counts conducted in the blue passband
showed a remarkable excess of faint galaxies relative to model predictions. 
This excess is known as the faint blue galaxy (FBG)
problem (see reviews by e.g. \cite{kk92}; \cite{e97}). Attempts to model the
counts and color-distributions of these faint blue galaxies led to the
conclusion that galaxies were more luminous and bluer in the recent past. In
order to better understand the field galaxy population, faint galaxy redshift
surveys were conducted which showed that standard luminosity evolution alone
cannot account for the excess of FBG's (e.g. \cite{bes88}). Many of the faint
galaxy spectra show evidence for strong star-formation, which led to the
conclusion that the steep slope of the number counts is produced by lower
luminosity galaxies undergoing short bursts of star-formation (\cite{bes88}).

More recently, many ground-based redshift surveys have been conducted to faint
limits in order to further study this issue. The Canada-France Redshift Survey
(\cite{cfrs95a}; hereafter CFRS) contains 591 galaxies brighter than
$I_{AB}\lesssim22.5$ mag. They have studied the redshifts, emission line
strengths and ground based photometric properties. With this data, Lilly \etal
(1995b) showed that the evolution in the luminosity function (LF) out to 
$z<1$ was greater for the bluer galaxies than for the redder ones (presumed to 
represent late and early morphological types, respectively). Recently, this
work has been extended to include HST based morphology (\cite{cfrs1};
\cite{cfrs2}; \cite{cfrs3}; \cite{cfrs4}), as we will use for a larger number of
field galaxies in this paper. The Canadian Network for Observational Cosmology
cluster redshift survey (\cite{cnoc1}) aimed at studying galaxy clusters
in the range $0.2<z<0.55$, and was complete for {\it field}
galaxies with $z<0.3$. This work has been extended to include more
galaxies (\cite{cnoc2}). For a better measurement of the local galaxy
luminosity function, the 2dF Galaxy Redshift Survey will provide spectra
and redshifts for 250,000 galaxies to a limit of $b_J=19.45$ mag 
(\cite{2df}; herafter 2dFGRS). Within several years, the Sloan Digital 
Sky Survey will provide redshifts for $\sim 2 \times 10^{6}$ field galaxies 
to r'=18.0 mag or $\Bj \lesssim 19$ mag (\cite{gun98}; \cite{blan01}; 
hereafter SDSS). These surveys provide very significant 
spectroscopic coverage of relatively nearby galaxies. However, as we will 
show in this paper, the median scale-lengths of galaxies at $\Bj \gtrsim 19$
mag is $r_e \lesssim 1\farcs 0$, and rapidly decreases at fainter magnitudes,
so that reliable morphological information 
for faint galaxies ($\Bj\gtrsim 19-20$ mag) over wide fields is beyond the 
capabilities of ground-based facilities and has to be done from space. 

The advent of high resolution space-based optical imaging opened the door for
studying the sub-arcsecond properties of many types of astronomical objects
(e.g. \cite{d95a}; \cite{glaz95}; \cite{a96a}; \cite{o96}; etc). In
particular, the Hubble Space Telescope (HST) has proven very useful for
studying the properties of faint, as well as distant galaxies. Most notably the
Northern Hubble Deep Field (HDF--N, \cite{wil96}) provided deep ($I\lesssim29$
mag) imaging of a single field in the $UBVI$ filters at a resolution of better
than $0\farcs1$ (actually $\sim0\farcs06$ FWHM after {\it drizzling}). There
have been many studies of faint galaxy morphology from this data set ({\it e.g.}
\cite{o96}; \cite{a96b}; \cite{vdb96}; \cite{d98}; \cite{ms98}). The HST Medium
Deep Survey (MDS) covered a much larger area of sky, but to a lesser depth,
and took advantage of HST's parallel observing mode to image many fields in 
mostly $V$ and $I$ (e.g. \cite{rgo99}). This large data set has produced many
studies (e.g. \cite{dwg95}; \cite{a96a}; \cite{im96}; \cite{roc96}; 
\cite{im99}), including one that utilized the small fraction of the MDS data
that was observed in the $B$--band (\cite{roc97}).

The HST data allowed for the morphological properties of these faint field
galaxies to be studied. One of the main discoveries was that most of the FBG
excess is due to galaxies of irregular or peculiar types (\cite{d95a};
\cite{dwg95}; \cite{glaz95}). These authors showed that the counts are in
excess of the {\it non-evolving} LF model predictions, regardless of adopted
cosmology, with the caveat that the local luminosity function has to be
(arbitrarily) re-normalized by a factor of up to 2 at a flux level of $b_j=18$
mag. The justification for this re-normalization is typically given as
either due to: (1) local inhomogeneity (\cite{marz94}; \cite{zuc97});
(2) photometric scale errors (\cite{met95}); (3) rapid recent
evolution in the galaxy population (\cite{mad90}); and/or (4) the probability
that local surveys are biased against late-type low-SB galaxies in selection for
spectroscopic follow-up. Nevertheless, it remains an unsatisfactory situation 
that an ad hoc correction is required to reconcile the local measures of the 
luminosity function of galaxies with galaxy counts as bright as $b_J=19$ mags.
For more details on this issue, see discussions in Driver, Windhorst \& 
Griffiths (1995) and Marzke \etal (1998). It is therefore crucial to have 
good statistics, {\it with morphology} at this brighter flux level in 
the blue filter to attempt to explain, or perhaps even rule out, this 
apparently necessary normalization factor.

Further work on the evolutionary model predictions for faint field galaxies
has produced conflicting results. Without being able to fully cover the 
literature on this topic here, we will try to present the flavor of the
problem. The bright HDF--N galaxies were used to model what the distribution 
and appearance of the fainter HDF--N galaxies should look like assuming 
no-evolution (\cite{bbs98a}), as well as assuming different types/amounts 
of evolution (\cite{bbs98b}). Assuming that the 30 brightest HDF--N galaxies 
are a fair representation of the thousands of fainter ones, they showed that 
the no-evolution models could not simulate the observed properties of the 
population of the fainter galaxies (\cite{bbs98a}). It was then shown that pure 
luminosity evolution (PLE) models fit the galaxy counts, but not the galaxy size
distribution, and that dwarf augmented populations would not fit the counts 
(see also \cite{babul96}; \cite{fb98}). A model that does fit the data was that 
of ``mass-conserving'' density evolution (\cite{bbs98b}). These results should
be viewed with caution, since the HDF--N was selected to be devoid of bright 
objects (\cite{wil96}), which, compounded by the fact that galaxies are known to
cluster, would logically lead to a general deficit of low- to 
intermediate-redshift galaxies in this field. Kauffmann, Charlot $\&$ White 
(1996) used the ground-based CFRS colors and redshifts to predict which 
galaxies are of early-types, and then showed that more than the standard amount
of passive evolution was required by the data. It should be noted that the use
of colors to determine morphology of distant galaxies was shown to be 
problematic within the same data set, when the HST morphologies were added, 
and no evolution of the space density of ellipticals was observed 
(\cite{cfrs3}).  Another study showed that in an open universe, the counts 
(in the $U b_J r_k I K$ bands) and color and redshift distributions could 
be reasonably explained by pure luminosity evolution, but that number and 
luminosity evolution (NLE) would be required in an $\Omega = 1$ universe 
(\cite{poz96}). In contrast, it was shown that PLE models that fit the 
number counts and redshift distributions cannot fit the
observed $(B-K)$ color distribution regardless of chosen cosmology
(\cite{he98}). This was due to the fact that all galaxies with $(B-K) > 5.5$
mag must be ellipticals, and the PLE models do not match the observed redshift
distribution for these red galaxies. These authors showed that number luminosity
evolution better matched all the data that was considered (\cite{he98}). The
luminosity functions of elliptical galaxies out to $z<1.2$, with morphologies
determined from the HST-MDS, were constructed and showed $1 \pm 0.5 $ mag of
luminosity evolution, and argued against strong number density evolution
(\cite{im96}). Therefore, it is reasonable to say that the issue of the amount
or type (luminosity, density or some combination thereof) of evolution is not a
solved problem, and is further compounded by our poor understanding of the
LF normalization factor mentioned above.

Traditionally, galaxies have been classified visually by eye. The large
number of galaxies that are observed today necessitated the development of
automated, computer-based classification methods. Odewahn \etal (1996) used
artificial neural network (ANN) classification in a seven-dimensional 
photometric parameter space to estimate galaxy stage values in the revised
Hubble classification system (\cite{dev59}). Abraham \etal (1996a) used linear
relationships between two parameters, asymmetry and central concentration.
Marleau and Simard (1998) used a 2-dimensional bulge-disk decomposition to
compare the ratio of bulge-to-total light to the types of Abraham \etal (1996a)
and of van den Bergh \etal (1996). Though the results for a given galaxy may
differ for any of these methods, the overall qualitative results are in
reasonably good agreement.

Where do all these faint galaxy studies leave us? Currently, these HST surveys
have yielded rather poor statistics for galaxy morphology in the magnitude
range $\Bj\simeq18-24$ mag. Since the correspnding median scale-lengths of faint
galaxies are $r_e \simeq 1\farcs0 - 0\farcs5$ in this magnitude range
(see \S 4.2 and figures below), we cannot get reliable galaxy morphology 
$\Bj\simeq18 - 24$ mag over wide fields from the ground. Hence, there 
remains a need for better statistics {\it and} good HST morphologies in the
magnitude range of $18\lesssim b_J \lesssim24$ mag, where the earlier surveys
have provided little statistics. Since the number of galaxies observed at a
given magnitude is roughly proportional to the area surveyed, and since there
are more faint galaxies than bright ones, ultra-deep surveys like the HDF--N and
the field surrounding the weak radio galaxy 53W002 (\cite{o96}) have provided
important information on galaxies fainter than $\Bj\gtrsim 23$ mag. However, the
statistics get increasingly sparse for brighter galaxies. The goals, therefore,
of this paper are to survey a large number of HST fields in the $B$ and $I$
filters over a wide area to a lesser depth, and to study the properties of
these brighter galaxies as a function of observed $B$--band brightness. This 
will help to fill in this new, but essential portion of parameter
space and aid in the understanding of the faintest galaxies observed today.

The HST observations are described in \S~\ref{obs} and the data reduction and
catalog generation are outlined in \S~\ref{reduc}. The results of this paper
are presented in \S~\ref{results} and discussed in \S~\ref{discuss} in terms of
other recent work in the field. We summarize the paper in \S~\ref{summary}.
For clarification of nomenclature, we refer to the observed radius of a galaxy
containing half of the observed light as the effective radius ($r_e$), and this
is discussed in \S~\ref{magre}. Since the two deeper fields, HDF--N and 53W002,
are essentially complete past our classification limit at $\Bj\lesssim27$ mag
(\cite{o96}), we shall group them together and refer to them as the $B$--band
Deep Survey (BBDS), while the 31 shallower fields will be referred to as the
$B$--band Parallel Survey (BBPS). We adopt $H_{o} = 65\ km\ s^{-1}\ Mpc^{-1}$
and a flat $(\Omega_{M}=0.3,\Omega_{\Lambda}=0.7)$ cosmology throughout, 
except where otherwise noted.

\section{Observations} \label{obs}

All data for this survey were taken with the Hubble Space Telescope using the
Wide-Field and Planetary Camera 2 (WFPC2) in parallel mode during Cycles 6--7. 
Fields were randomly selected with the criteria that they be at high Galactic
latitude ($\mid b^{II}\mid \ \geq30\arcdeg$) and that they contain no bright
SAO stars, RC3 galaxies or known Abell galaxy clusters. The data set consists
of 31 fields, all selected to have low Galactic extinction, $A_B \le 0.4$ mag
(\cite{bh82}). The data are summarized in Table~\ref{tbl-1}. Note that fields
bb025 and bb026 were not used in the final analysis because they contained
single $I$--band exposures that were inadequate for this survey.

\placetable{tbl-1}

For each field, images were taken in both F450W (the WFPC2 $B$--band) and
F814W (the WFPC2 $I$--band) with a longer exposure time in $B$ due to the
approximate 20\% lower sensitivity at that wavelength (see Table 8 of
\cite{holtz95}). Combined with the expected relative colors of the faint field
galaxies compared to the Zodiacal sky, the total survey efficiency in F814W is 
$\gtrsim2\times$ better than in F450W. Therefore, we took 2--4 orbits in $B$
and two orbits in $I$. The reason for using both
filters is to select the galaxies from the $B$--band images, where the HST
resolution improves deblending over ground-based data, and therefore provides
the most reliable object selection and photometry. The $I$--band images 
(which corresponds to the rest-frame $B$-band) have significantly
higher S/N, and therefore provide the best possible source of morphological 
classifications, as well as provide color information
over a wider color baseline than the standard $V$--$I$ color (F606W--F814W)
that has traditionally been used in most WFPC2 studies. The two sets of images
coupled with a future redshift survey will also allow us to do  a study of the 
rest-frame color gradients in the brighter galaxies. The total area of the
BBPS survey, including the PC exposures, is approximately 0.0370 square
degrees. The WF(PC) data have a $0\farcs0996$($0\farcs0455$) spatial sampling 
and a limiting magnitude of $\Bj\lesssim23.5$ mag (the approximate 90$\%$ 
completeness limit for compact objects), as discussed in \S~\ref{morphcts}.

\section{Data Reduction} \label{reduc}

\subsection{Image Stacking}

All images were registered using the centroids of manually selected compact and
bright objects, with the centroids being determined from the IRAF task {\it
imcentroid}. Since the detectors are fixed relative to each other, shifts were
determined for each of the 4 WFPC2 CCDs, and then averaged together to get the
optimal offset for each set of exposures in a given field. These offsets were
rounded off to whole numbers and only integer shifts were applied, so as not to
introduce additional numerical noise into the data (\cf \cite{rgo99}). Due to
the nature of parallel observing with HST, a set dither pattern cannot be
chosen by the parallel observers, and so the method employed here produces
the most self-consistent data set. The appropriate shifts were then applied 
taking into account the different chip orientations and the finer sampling of
the PC.

Since most fields were two orbit cases, it was necessary to improve upon
existing cosmic ray (CR) rejection algorithms, which were optimized for use
with $N\geq5$ images (e.g. \cite {wfn94}). All image stacks were created using
a customized IDL routine which was specially developed for this project. This 
routine was also used for the low light-level images in the F410M filter of
Pascarelle, Windhorst $\&$ Keel (1998). Images through different filters
were handled separately to ensure that our final morphological 
classifications were as color-independent as possible. The reason for
developing this new routine was the need to accurately reject CRs over a few
(2--4) independent exposures, while assuring at high reliability that the
science image itself would not be corrupted by the algorithm. 

For $N$ registered images, the IDL routine performs the following operations at
{\it each} pixel (x,y) location {\it separately}. It creates a list of $N$ pixel
values which is then sorted from the lowest to the highest value. The following
Poisson noise model based on the known CCD characteristics is then used to
determine which pixel values should and should not be included in the average:
\begin{equation}
\sigma_{x,y}=\sqrt{DN_{x,y}*g+RN^{2}+DK*t}/g
\end{equation}
where $DN_{x,y}$ is the number of ADU in pixel (x,y), $g=7.0$ $e^{-}/ADU$ is
the WFPC2 detector gain, $RN=5.3 e^{-}$ is the read-noise, $DK=0.0033 e^{-}/sec$
is the dark current rate and $t$ is the exposure time in seconds.  Starting with
the {\it lowest} pixel value, each successive value is checked to see if it is
within 2.5 $\sigma$ of the current average value. This way, higher pixel values
that are likely due to cosmic rays are rejected (at the 2.5 $\sigma$ level).
This process is then repeated for each pixel. This rejection algorithm will
fail for the two orbit cases when a pixel is hit by a CR in {\it both} images,
which we measured to occur about 0.3\% of the time, or for about $\sim2000$
pixels that were affected by CR's in both full-orbit WFPC2 CCD exposures. This 
number is in excellent agreement with the values given in the {\it WFPC2 
Instrument Handbook}
\footnote{Updates to the WFPC2 Instrument Handbook can be found at:
http://www.stsci.edu/instruments/wfpc2/wfpc2\_top.html}
(\cite{bir00}) 
when extrapolated to our longer exposure times. These left-over CR's cause
problems with automated image extraction, because they would be counted as
objects and possibly classified as faint galaxies, which would
contaminate the sample.

In order to deal with this problem, we checked for cases of ``double hits'' by
subtracting a 3$\times$3 pixel median-smoothed image from each original image,
and compared this difference image to the difference image for the other
orbital exposures. This way, all ``double hits'' were located and interpolated
over by substituting the median of the 8 surrounding pixels. The result was a
much cleaner looking image, but this process also unavoidably generated holes
in the centers of bright, centrally-concentrated objects (stars and early-type
galaxies). Since these holes would obviously contaminate the photometry of the
bright galaxies we are interested in, we opted to use the cleaner but
non-photometric images {\it for the object selection only}. Once the object
positions were found, all analysis was performed on the images that still
contained the ``double hit'' CR's, but that had objects that were uncorruptedin their central regions.

\subsection{Object Extraction}

All image detection was done using the SExtractor version 1.0a (\cite {BA96})
object finding software package. Each WFPC2 WFC (or PC) image is convolved
with a 7$\times$7 (or 9$\times$9) convolution mask of a Gaussian kernel with a
FWHM of 3 (or 5) pixels. Then all objects with at least 8 contiguous pixels
that are 2.5 $\sigma$ above the local sky-background (as determined by
SExtractor) are extracted, and their location, size, and magnitude are written
to a file. The mean isophotal 2.5$\sigma$ detection limits for each field are
listed in Table~\ref{tbl-1}. Averaged over all fields, this surface brightness
(``SB'') detection limit corresponds to SB($\Bj$)$\simeq24.2\pm0.2$ 
($22.8\pm0.2$) \magarc for the WF (PC) detectors. The 1$\sigma$ SB sensitivity 
limits for 53W002 field of Windhorst, Pascarelle \& Keel (1998) are
SB($\Bj$)$\simeq$27.5\magarc and SB(I)$\simeq$26.7\magarc, and these numbers
are about $\sim$1.5 mag fainter for the HDF-N (\cite{o96}). 

For each detected object, a smaller sub-image is then cut whose size is 
determined by the SExtractor image parameters.  The subsequent
image analysis is done on these individual object images (i.e. all surface
photometry and morphological classification), using the MORPHO package of
Odewahn \etal (1996, 1997).

Once the sub-images are extracted, all objects are visually inspected to
determine whether they are true detections, as well as to determine if
SExtractor over-did the deblending of neighboring objects. The latter was
especially a problem for the brighter spiral and irregular galaxies that had
several peaks in their 2--dimensional brightness distributions, and is a
general problem for other object finding algorithms as well (see e.g. 
\cite{val82}; \cite{nw95}). The number of objects that had to be re-extracted
by hand was relatively small, about 10 per field. The number of separate
objects that were close enough to overlap was also small, but SExtractor
generally did an excellent job of deblending these, although we emphasize that
any deblending process is, by definition, somewhat arbitrary. The photometry
of these blended objects is addressed below.

\subsection{Surface Photometry}  \label{surf}

Photometric zero-points were taken from Tables 9 and 10 of Holtzman \etal
(1995). In the visual--red, these have not changed much during the lifetime of
WFPC2 (\cite{bir00}). We used the ``synthetic'' WFPC2 zero-points (Vega-based),
which means that the F450W magnitudes are equivalent to 
$b_J$\footnote{Kron (1980) gives the transformation from standard to
photographic $B$ magnitudes as $B-\Bj=0.23(B-V)$ and, using 
Holtzman \etal (1985) Table 10, we derive $B-F450W=0.23(B-V)$, which shows
that the synthetic WFPC2 F450W system is the same as the photographic $\Bj$
system.}, and the F814W magnitudes are the same as the ground-based $I$
magnitudes. This is to allow optimal comparison to previous ground-based
galaxy $B$-band counts, which were primarily done in the
$\Bj$ filter (or a filter that easily transforms to $\Bj$). For completeness,
we used zeropoints of $ZP_{F450W}=21.929$ mag and $ZP_{F814W}=21.676$ mag for 
$\Bj$ and I, respectively, for a count of 1 DN per second of exposure time.  

In order to accurately derive light-profiles and compute total magnitudes, a
good determination of the sky-background must be made. For each field and
filter, sky-values are determined for each of the four WFPC2 CCD's using the
following technique. The CCD is divided into 16 squares of 101$\times$101
pixels each. Next, a sky-value and sky-sigma are determined for each square by
fitting a ``tilted plane'' (\cf \cite {nw95}) to that area, after rejecting all
real objects that were found at the $\geq 2\sigma$ above the local sky level.
The reason for not fitting a more complicated surface is that the images are
sufficiently flat-fielded over $\sim100$ pixel scales, so that mostly linear
sky gradients are left, if any. Residual sky gradients could be due to, \eg
the telescope pointing too close to the Earth's limb, which is not always
controllable when scheduling parallel observations. This could produce a simple
mono-directional linear gradient in the WFPC2 sky, although this was not often
observed. The median of the 16 sky-box values was then used to determine the
final sky-value and sky-sigma. These values were found to agree with the
global {\it BACKGROUND} and {\it RMS} values determined by SExtractor to within
2$\%$. 

Since the Zodiacal sky at moderate to high Ecliptic latitudes is about
SB($\Bj$)$\approx$23.8--24.1\magarc and SB(I)$\approx$22.15--22.45\magarc
(\cite{w98}), a 2\% error in the sky-subtraction becomes the dominant 
factor at surface brightness levels SB($\Bj$)$\approx$28.0--28.3\magarc or
SB(I)$\approx$26.4--26.7\magarc, at which level this sky-subtraction error
equals $\sim$100\% of the galaxy SB-profile signal. In the noisier BBPS
images, we cannot push the galaxy SB-profiles this faint, but for the two BBDS
fields this is one reason why we do not push the galaxy classifications (see
\S 4.1) fainter than total fluxes of $\Bj=27.0$ mag (or $I=25.5$ mag). This is
because at these total flux levels, the median galaxy scale-lengths are
$r_e =0\farcs 2-0\farcs 3$ (see \S 4.2), so that the sky-subtraction errors in
the average SB-value out to \re are of order 10--20\%. This is the maximum
acceptable error in SB-profile parameters to allow reliable ANN
classifications, which are based on the galaxy SB-profile and other
parameters (see \S 3.5). Fainter than $\Bj \gtrsim 27$ mag, ANN galaxy
classifications are mainly unreliable due to a lack of a good training
set (\cite{o96}), which would be necessarily based on the abilities
to assign eyeball classifications as well as accurately measure the light
profiles of these galaxies, both of which become increasingly suspect
fainter than this flux level, for reasons stated above. The issue of uncertain
resdshifted UV-morphology could also play a role here and is further discussed
in \S~\ref{uvmorph}. Given that $b_J\simeq27.0$ mag is {\it also} the limit
due to the sky background subtraction (at these wavelengths), there is little
hope of reliable classification for fainter galaxies with HST, {\it even 
with improved resolution and/or deeper images}.  

All surface photometry was performed on each sub-image, using these
sky-values, with the MORPHO package as described in Odewahn \etal (1996,
1997). Since each sub-image to be analyzed has to contain a single object, the
analysis on crowded areas was done as follows. If two objects were too close
together, two sub-images were generated, each with {\it one} of the two objects
replaced by the local sky value, as interpolated from surrounding unaffected
sky areas. The size and shape of the ``blanked'' out region was an ellipse based
on the SExtractor values for position, size, axis ratio and position angle. 
This way, when the isophotal ellipses were fit to determine the
light-profile, the target object was minimally affected by its neighbor. Since
the total magnitude is determined from the integrated light-profile, this was
the most reliable way to handle proximate objects, while reducing the effects of
crowding on the other measured object parameters. Note that because the ANN
classification is based on individual object light-profile based parameters 
(\S~\ref{galclass}), and simultaneous light-profile fitting on {\it more} than
one object is very difficult to do (\cite{schm97}), this seemed to be the most 
robust way to get an accurate ANN class for every object, even if close
neighbors were present.

\subsection{Catalog Generation}

Once the images in both the $B$ and $I$ filters are processed and cataloged,
the $B$ and $I$ object lists are cross-matched by pixel location. For each
object in the $B$--band catalog, the $I$--band catalog is searched using a
search radius of 3 effective radii. The effective radius \re used here is the
half-light radius as measured from the $B$--band image. With this search
criteria, about 70\% of the objects in the $B$--band catalog have $I$--band
counterparts, while more than 95\% of the $I$--band catalog have $B$--band
counterparts. A manual check of the images showed that almost all the
unmatched $B$--band catalog objects were not real galaxies, but likely image 
defects, noise spikes or remaining cosmic rays left in the $B$--band images. 
We thresholded at $2.5\sigma$ so as not to exclude any real objects, at the
expense of making false detections which would get thrown out in cross-matching.
Since the $I$--band images are of higher quality and go deeper, despite their
generally shorter exposures, it is doubtful that very blue objects were excluded
through this selection process. However, this matching procedure could select
against extremely red objects, which we must bear in mind during the 
interpretation (\S 4 and 5). 

\subsection{Morphological Galaxy Classification} \label{galclass}

Galaxy types are assigned to all objects using an automated artificial 
neural network (ANN) pattern classifier that is based on the
{\it shape} of the object light-profile. Galaxy types, or hereafter
T--types, were designated on the 16 step revised Hubble system
(\cite {dev59}). This classification scheme was chosen because
T--types have been shown to correlate well with physical properties
in nearby galaxies (\cite{rh94}) and the available data used to
produce our galaxy count models are segregated in the same way. Since the
typical machine classification error in type is 2--3 steps (\cite{o96}),
types were rebinned into three broad intervals, that each span a much
larger range than the ANN classification error (E/S0, Sabc, Sd/Irr). This
classification error results from noise in the profile-parameters of the
faint objects (as discussed in \S~\ref{surf}), and from inconsistencies
between experienced classifiers, who visually classified a limited number
of galaxies in the ``training set'' that is used to train the ANN before it
is run on a much larger sample. To the extent possible, the personal offsets
and biases between these expert classifiers were removed (\cf \cite{o96};
\cite{O02}) before constructing the final training set, thereby reducing
systematics in the ANN classifications (although not yet proving that these
classifications are necessarily correct). The ANN's used
here are the rest-frame networks developed in Odewahn \etal (1996), which
were trained on the HDF--N galaxies that had published redshifts at that time.
The redshifts are required to know the rest-frame wavelength in which 
each galaxy in the $UBVI$ training set was actually observed.
T--types are in the range $-7\leq T < -0.5$ for objects
designated as ``E/S0,'' $-0.5\leq T < 5.5$ as ``Sabc'', and $5.5\leq T < 12$
as ``Sd/Irr''. 

We note that the ANN's of Odewahn \etal (1996) always produces an answer, 
\ie based on the
actual galaxy SB-profile parameters, it forces a classification in one of these
three T-type intervals, no matter how faint or how low-SB the object is. That
is, we deliberately sacrifice classification accuracy to get 100\% completeness
in the classifications, so that no objects remain unclassified. The corollary
of this is, of course, that beyond a certain flux and SB limit, the
classifications become increasingly unreliable due to the lack of an
adequate training set. The lack of an adequate training set at these
faint flux levels is due to a combination of insufficient S/N and resolution
for a human classifier to assign a reliable type, as well as a poor 
understanding of the rest-frame UV morphology for the higher redshift objects
where this is a potential issue. As discussed in \S 3.3, we
believe that this limit occurs in the two BBDS fields at total fluxes of
approximately $\Bj=27.0$ mag ($I=25.5$ mag). We will show in \S 4.1 that the
shallower BBPS fields produce consistent classifications close to their
completeness limit of $\Bj\simeq23.5$ mag in the flux range of overlap with the
two BBDS fields ($22.5\lesssim\Bj\lesssim 24.0$ mag), where the 
classification of the latter objects are robust, since they have much
higher S/N at the same flux level. 

Using the suitably trained neural networks, galaxy types were generated from
both the $B$--band and $I$--band images, as described below. However, for a
variety of reasons, all types used in this paper are based on those ANN-values
assigned from the $I$--band images. First, the $I$--band images are superior
to the $B$--band ones due to the larger overall sensitivity and the much darker
Zodiacal sky background at that wavelength, even for the bluest detectable
field galaxies. The second reason is bandpass shifting. The median redshift of
the fainter galaxies at $\Bj\leq25$ mag is $z \simeq 0.6-0.7$ (\cite{kk92};
\cite{e97}). Therefore, the light observed in the $I$--band was emitted in
approximately the $B$--band where morphology of local galaxies has been well
studied, and a good training set for the ANN exists. Analogously, the light
observed in the $B$--band is emitted in the rest frame near-UV, where 
systematic studies of nearby objects have only recently been done (\cf
\cite{glb96}; \cite{bwo97}; \cite{kuch01}; \cite{marc01}; \cite{win02}). 
The issue of uncertain rest-frame UV morphology and its effect on the galaxy
counts as a function of type is further discussed in \S~\ref{uvmorph}. 
For the BBPS, the use of $I$--band classifications allows us to take advantage
of bandpass shifting instead of being hampered by it. Thirdly, the effects of
the k-corrections are smaller at this longer wavelength. Finally, the HST 
images are closer to being properly sampled in the $I$--band, compared to the
$B$--band, allowing for better classifications.

We classified all these galaxies from their $I$--band images, but using an ANN
defined at a shorter wavelength which is closest to the rest-frame wavelength
sampled by the $I$-band for that particular object. These rest-frame ANN's
were constructed from the HST $UBVI$ images in the HDF--N for objects with
spectroscopic redshifts ($z\lesssim1$). The details of this classification
method and the use of rest-frame ANN's are given in Odewahn \etal (1996, 1997)
and references therein. For this project, seven parameters are used as input to
the ANN which then produces a type. These parameters are all based on the
object's azimuthally averaged radial surface- brightness profile and are:

\begin{enumerate}
\item{the SB at the radius containing 25\% of the total light;}
\item{the SB at the radius containing 75\% of the total light;}
\item{the mean SB within the effective (half-light) radius;}
\item{the mean SB within an isophotal radius ($SB(I)=24.5 mag \ arcsec^{-2}$);}
\item{the slope of a linear fit to the profile in $r^{1/4}$ space;}
\item{the intercept of a linear fit to the profile in $r^{1/4}$ space; and}
\item{the axial ratio (b/a) of the outer isophote.}
\end{enumerate}

It is important to note that in the ANN classifications, color is not used
{\it at all} and that the effective radius is not used {\it directly}.
Fig.~\ref{imagre} shows that this indirect use of effective radius does result
in a loose correlation of $r_e$ with type in any given $I$--band magnitude
slice. It is also important to reiterate
that no bulge-to-disk decomposition or model profile-fitting is used. This
particular choice of the 7 ANN parameters is somewhat arbitrary. Different
sets of structural parameters were tested (\cite{o96}) and produced similar
classifications to within the errors quoted above.

As discussed in Odewahn \etal (1996), the classification limit in the two BBDS
fields is $\Bj\lesssim 27$ mag (or $I\lesssim 25.5$ mag).  At fainter flux
levels, human classification becomes increasingly unreliable. As illustrated in
\S~\ref{results} (Figures 3--7), it is interesting to notice that the ANN
classifications appear reasonable, even a magnitude or so fainter than the
visual classification limit of the training sets.

These classifications are reasonable because the parameters that are not used
in the classifications (color and effective radius) follow the type-dependent
trends for galaxies at cosmological distances that one would expect from the
known properties of local galaxies (see discussions in \S~\ref{magre} and
\S~\ref{cmd}). However, this is not proof that these classifications are
therefore correct, and two things are required to verify these classifications
beyond the currently stated limits:
higher resolution images of these galaxies and a better mid-UV training set for
the ANN (see \S~\ref{uvmorph}). For the brighter BBPS sample, the $B$--band
data is too noisy near
its completeness limit to believe the classifications from the $B$--band images
alone. This is another reason for classifying in the $I$--band images, which go
$\approx0.5-1.0$ magnitudes deeper even for blue galaxies, and therefore have a
higher signal-to-noise ratio. Therefore, all BBPS fields have reliable
classifications to their respective $B$--band detection limits listed in
Table~\ref{tbl-1}, and the two BBDS fields have reliable classifications to
$\Bj\lesssim27$ mag (ie. $I\lesssim25$ mag). 

\subsection{Star-Galaxy Separation} \label{sgsep}

The issue of star-galaxy separation is an important one, especially in a
survey of this nature where the goal is to count different types of objects,
including the most compact ones. The star-galaxy separation for the BBPS was
initially performed using a linear division in a single 2--D photometric
parameter space. The effective radius, $r_e$ forms the Y axis. The X axis is
formed with the ($MORPHO$) ``C42'' parameter. This is a concentration index
formed by the ratio of the size of the ``total aperture'' to the 50\% flux
aperture (i.e. the number of effective radii inside the total aperture). Both
apertures have an elliptical shape defined by the median ellipticity of the
ellipse fits to the lower SB isophotes. The determination of the ``total
aperture'' is a complex procedure, which finds the optimally sized aperture
based on the local signal-to-noise ratio of the sky. An object is classified as
a star if $C42\geq60.0$, or if it lies below the line $r_{e}=0.010*C42-0.095$.
These values were determined using the F814W data from the HDF--N, the HDF--N
flanking fields, the 53W002 field and the 6 earlier MDS fields from Driver,
Windhorst \& Griffiths (1995). They were plotted for several magnitude cuts
using symbols to indicate the morphological types assigned in our past visual
classification
work. Bright objects with clear diffraction spikes were always visually classed
as stars. These could erroneously include a few QSO's, if these were not
otherwise recognized, \ie from their available spectra (\eg \cite{p96}) or weak
radio fluxes (\cite{w98}). By inspecting many such plots, the best line that
divided the stars and the E/S0 types was determined. At bright magnitudes
(usually $I\leq22$ mag for most BBPS fields), the parameter space always showed
a fairly substantial segregation between stars and even compact, high SB
galaxies (likely E/S0). Of course, as we go fainter, this division blurs due to
increasing noise in the extracted $r_{e}$ and C42 measurements. This separation
limit is similar to the one found by M\'{e}ndez $\&$ Guzm\'{a}n (1998), who
used a different method. Also, a visual inspection of the 40 objects with
$\Bj<21$ mag and not classified as stars showed that this method failed for 12
bright ($\Bj<19$ mag) and saturated stars. These 12 stars were removed from the
the final galaxy counts. 

\begin{figure}
\plotone{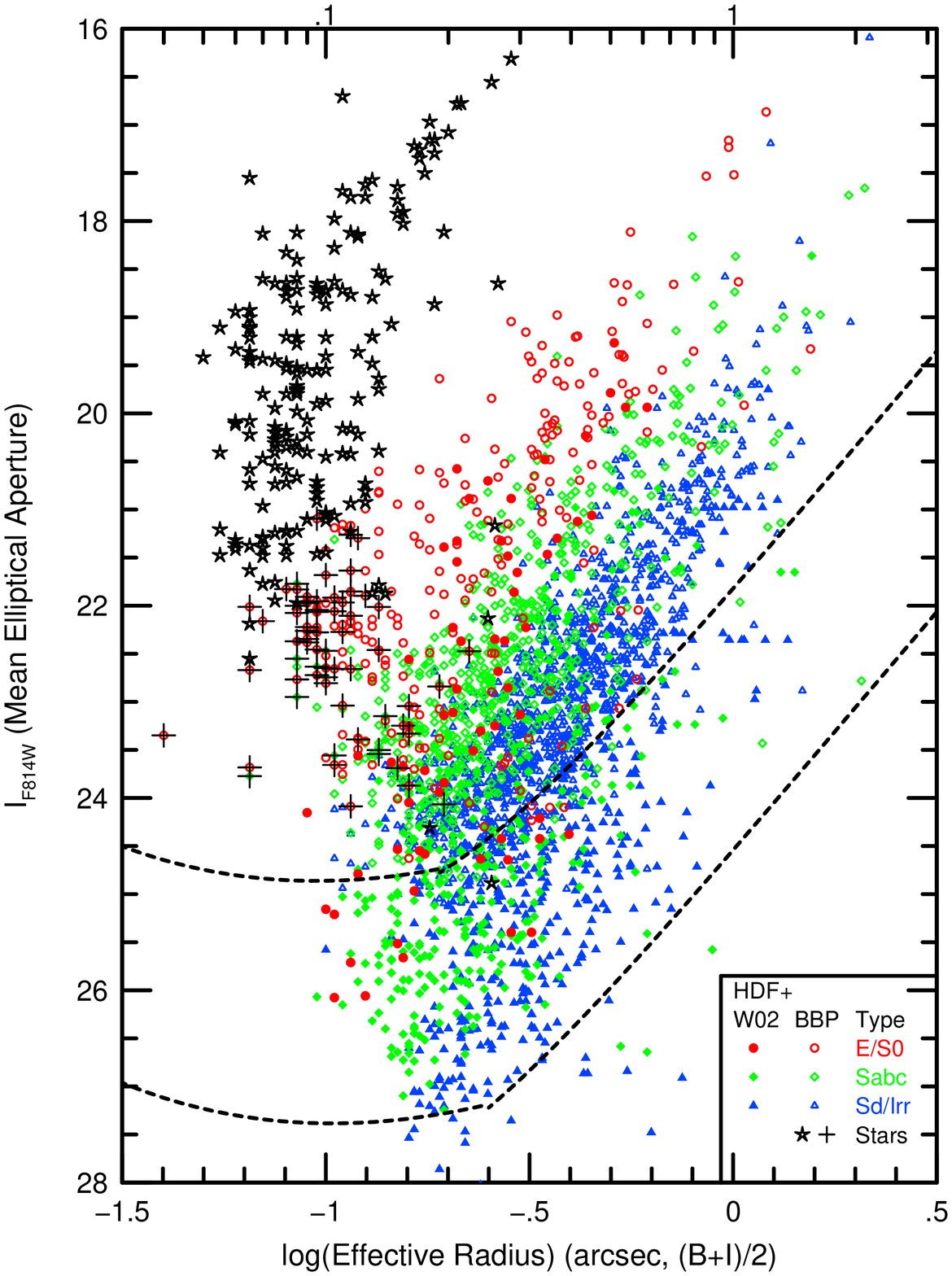}
\figcaption[SHCohen.fig1.ps]{$I$--band magnitude vs. effective radius for the
BBPS. This figure shows that misclassification between stars and galaxies
occurs for $I\gtrsim22$ mag. At this magnitude, a fraction of stars may be
classified as early-type galaxies. Note that stars are only plotted for the
BBPS fields and not for the two BBDS fields. The dashed lines indicate the
approximate completeness limits for the BBPS and BBDS, as discussed in the
text. The objects with pluses superposed were re-classified as ``stars'', and
are likely globular clusters in the Virgo cluster, as explained in
\S~\ref{sgsep}. The scale indicated across the top is for $r_e$ in arcseconds.
 \label{imagre}}
\end{figure}

To demonstrate the success of this star-galaxy separation method,
Fig.~\ref{imagre} shows the $I$--band magnitude versus effective radius
relation for the BBPS objects, including those objects that were classified as
stars. It is clear that to the BBPS $I$--band compact-object detection limit
($I\simeq23.75$ mag), the stars easily separate out from most of the galaxies,
except for an interesting group of objects, most of which were initially
classified as ``elliptical'' galaxies, that are small 
(log($r_{e}\arcsec) \simeq -1$) and near the faint end of the sample
($I\simeq22.0$ mag -- these objects are plotted as pluses in Fig. 1). This
raises the obvious and important question of whether or not these objects are
misclassified stars.

\placefigure{imagre}

\begin{figure}
\epsfig{file=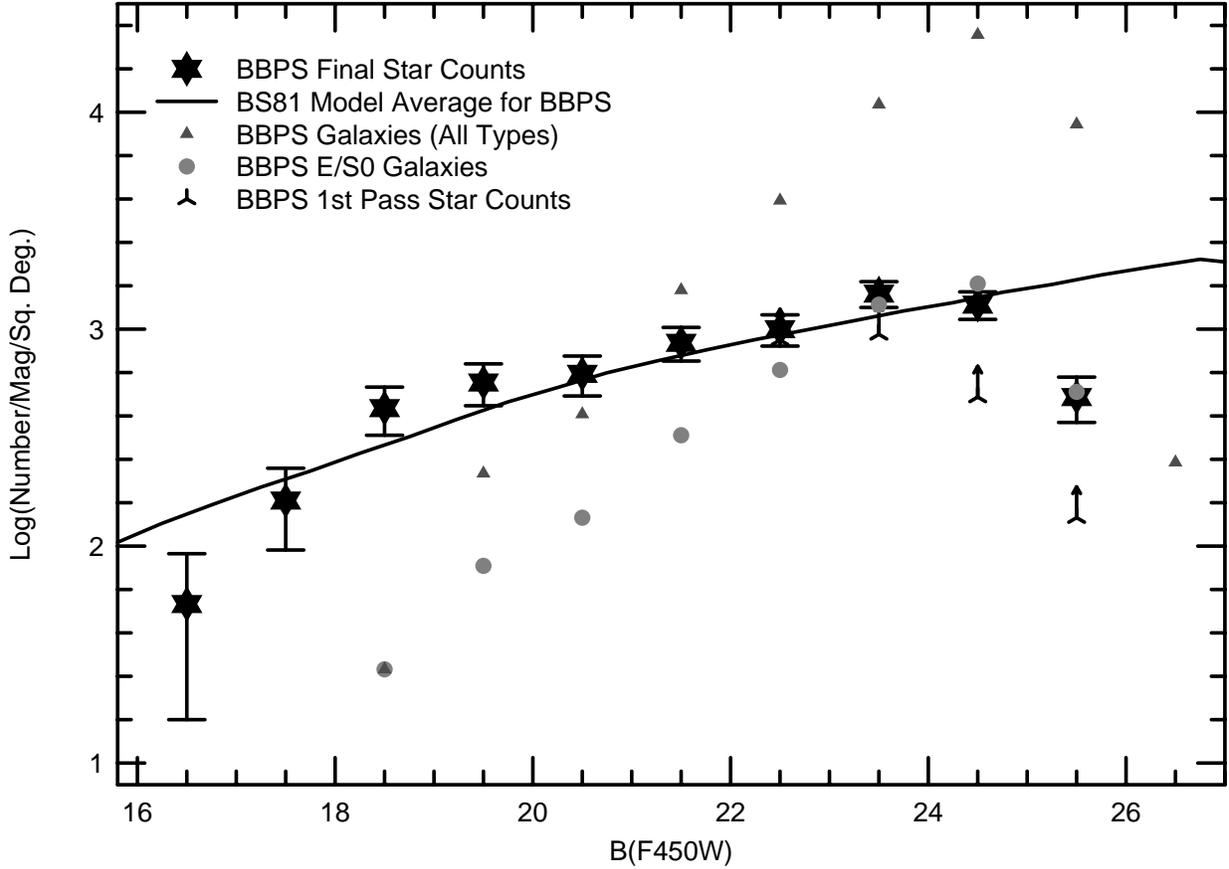,width=7.5in}
\figcaption[SHCohen.fig2.ps]{Differential star counts for the BBPS data. 
Objects initially classified as stars are shown as upward pointing arrows and
the final star counts are shown as stars with error-bars (see \S~\ref{sgsep}). 
The solid line shows the predicted star counts from the Bahcall $\&$ Soneira
(1981) Galaxy model averaged over the Galactic coordinates of all 29
different HST parallel pointings. For comparison, the total counts for
objects classified as galaxies (triangles) and those classified as E/S0's
(circles) are also shown (see also Fig. 3). Note that to $b_J\lesssim23.0$ mag,
the initial star-galaxy separation is good.  Beyond this limit, the galaxies
begin to far outnumber the stars, so that a few misclassifications of stars
will not substantially affect the results on faint galaxies. However, at
$\Bj \gtrsim 23$ mag, the star counts become less reliable.
\label{starcounts}}
\end{figure}

In order to test this issue, the Galaxy Model\footnote{Fortran code to compute
Bahcall \& Soneira Galaxy Model is available at 
http:$\slash\slash$www.ias.edu$\slash$$\sim$jnb} of Bahcall \& Soneira (1981) 
was
used to predict the star-counts in these fields. The model was run for the
($l^{II}$,$b^{II}$) coordinates of $each$ of the 29 telescope pointings and the
resulting predicted star counts were {\it averaged} together. 
Fig.~\ref{starcounts} shows the differential number counts in the $B$--band for
all non-stellar objects, for the objects classified as stars, for the E/S0
galaxies only, as well as for the average Bahcall and Soneira model. The first
issue to notice is that the average star count model matches the observed star
counts very well. When compared to the star count model, it is clear that our
first pass of the star counts (plotted as upwards arrows in Fig. 1) turn over
at $\Bj\geq23.5$ mag due to a significant fraction of misclassifications and 
incompleteness, even before the galaxy counts noticeably turnover at
$\Bj\geq24.5$ mag. It
is therefore highly probable that some fainter stars have been misclassified as
galaxies. These would have most likely been misclassified as centrally
concentrated (and therefore elliptical) galaxies, as seen in 
Fig.~\ref{imagre}. There is also a color selection effect, in that the 
fainter stars are expected to have redder colors and therefore may have 
escaped the blue selection limit. Note that, in order for an object to appear 
in our catalogs, and hence in Figs. 1--7, it had to be seen in {\it both} 
the $B$ and $I$ filters. The majority of the stars from the two deeper fields 
with $I\geq22$ mag are too red [$(B-I)\geq4$ mag ] to have been selected in 
the BBPS sample. This is less of a problem for the galaxies because, 
as Fig.~\ref{cm} shows, at $I\geq22$ mag galaxies have ($B$--$I$) colors 
in the range of 1.0--2.5 mag. Fortunately, at this
brightness level and at fainter flux levels, the galaxy counts are much higher
than the predicted star counts (see Fig.~\ref{starcounts}), even for the
early-type galaxies which have the shallowest counts slope (see also
Fig.~\ref{galcounts}). Therefore, the likely contamination of the faint galaxy
counts by misclassified stars is minimal and not significant. The few 
missing bright ($\Bj\lesssim 18$
mag) stars in Fig.~\ref{starcounts} are due to the saturation problem mentioned
above. These missing bright stars were obviously detected, but due to
saturation, no reliable fluxes could be measured, so they are not plotted in
Fig.~\ref{starcounts}.

\placefigure{starcounts}

Since the star-counts should go as deep as the total object counts, and
Fig.~\ref{starcounts} shows that our first-pass star counts (shown as upwards
arrows) turns over before the galaxy counts, there must be some number of
fainter stars that were not classified as such. Therefore, a second step in
star-galaxy separation was applied to rectify this situation. The model star
counts were used as a reference point to update the stellar classification for
$\Bj>23$ mag. {\it Therefore, the resulting star counts can not and must not be
used fainter than $\Bj>23$ mag to test any star count models.} This is not a
problem here, since the goal of this study is to properly count the
{\it galaxies} in order to compare to galaxy count models as a function of
galaxy type.  A diagonal line was drawn in the magnitude--log($r_e$) parameter
space, where both quantities are measured in the $I$-band. Note that
Fig.~\ref{imagre} shows the effective radius ($r_e$) as an average over the
values measured in the $B$ and $I$ bands, while all star-galaxy separation
(and galaxy classification) was done exclusively on the $I$-band images. 
All objects to the left of this line were declared stars and all objects to 
the right were declared galaxies. This step did not change the classification
of any previously determined stars. The process was iterated by trial and error
until the observed star counts matched the predicted star counts up to the 
completeness limit of the total counts. The final star-galaxy separation line
that was used is given by $I=-20.313\times\log(r_e)+6.188$. The
objects that were re-classified as stars are shown as superposed pluses in
Fig.~\ref{imagre}, and it can be seen that the majority of them were originally
classified as ellipticals.

However, for any reasonable application of this second pass at star-galaxy
separation, the new star counts were always several sigma above the Bahcall
and Soneira models (at the faint end) and so warranted further investigation.
The colors of these faint centrally concentrated objects were examined and
a significant number of them were grouped
around $\langle(B-I)\rangle=1.44\pm0.42$ mag and the majority of these came
from 2 HST parallel fields in the Virgo Cluster, bb016 and bb018. Using the
NASA/IPAC Extragalactic Database (NED), it was realized that field bb016 is only
$8\farcm2$ away from M87. This well studied elliptical galaxy is known to have
a large and extended globular cluster population (\cite{H86}). If we designate
all the compact objects in this field with the colors mentioned above as
globular clusters in Virgo (loosely associated with M87), and compare the
number that we find (N=19) to the surface density of these objects in Harris
(1986), we get very reasonable agreement. The apparent magnitude of these
objects, combined with the M87 distance, also argues that these have the
expected luminosities of Virgo globular clusters. Inspection of the location of
the bb018 field shows that it is situated between two bright Virgo ellipticals
M84 ($8\farcm2$ away) and M86 ($9\farcm9$ away). It is therefore highly likely
that the bb018 star-like objects (N=18) in this color range are globular
clusters in the (overlapping) halos of both of these galaxies. The final
corrected star counts, shown in Fig.~\ref{starcounts}, have these globular
cluster candidates removed, and show that the updated star-galaxy separation
has no noticeable effect on the total galaxy counts, but do impact on the
counts of the faintest ellipticals, albeit minimally. In fact, given what is 
known about faint star counts (\cite{fgb96}), even if all faint stars were 
mis-classified as early-type galaxies, the E/S0 counts would be {\it at most} 
10\% too high in this flux range.

We emphasize that star-galaxy separation with HST is non-trivial, even at the
relatively bright limit of $I\approx 22$ mag, because faint galaxies generally
have small angular sizes (see \S 4.2 and Fig.~\ref{hist15} below), and because
objects other than stars and galaxies can sneak into the HST samples. 

\subsection{Galaxy Counts as a Function of Type}

Objects classified as galaxies were counted in half-magnitude wide bins. The
counts for each field were combined in the following way. The completeness
limits in each field is different for both observational (mainly varying
exposure times) and statistical reasons (\ie cosmic variance). First, 
the completeness limit for each field is determined from the flux level at
which the total counts in that field turn over (cf.  \cite{nw95}). All counts
beyond the completeness limit are given a weight of zero while all counts
brighter than this limit are given a weight of one.  Therefore, no field
contributes to any magnitude  bin (in the combined counts) in which it
is less than 90\% complete. The center of the faintest bin that was used from
each field is listed in the last column of Table ~\ref{tbl-1}. This ensures
that the fainter points are properly handled given the different exposure
times. These weights are then used to keep track of the area that contributes
to the counts in each bin. This results in some of the fainter bins having a
smaller effective area than the brighter bins, as  indicated in Table 
~\ref{tbl2}. This is not a problem because there are many more faint objects
and hence the statistics in the faint bins are not compromised. Below we 
discuss two possible sources of contamination, other than the stars
already mentioned in \S~\ref{sgsep}.

\subsubsection{Local Large Scale Structure}\label{LLSS}

Some of the parallel fields had coordinates that placed them in or near the
edge of the Coma or Virgo superclusters (as indicated in Table ~\ref{tbl-1}).
In order to test whether there was any excess of bright galaxies from these
superclusters, the counts were compared for the cluster and non-cluster fields.
For fainter flux levels, the counts from the cluster fields were within the 
$3\sigma$ field-to-field variance of the mean of the non-cluster field counts.
The only statistically significant contaminants were two galaxies from 
bb019 ($\Bj=17.08, 17.22$ mag) and one from bb001 ($\Bj=17.97$ mag), which 
are in the Coma supercluster. This was the {\it primary} HST target which may,
therefore, have biased the bright galaxy sample in the WFPC2 parallels 
we are studying here. This a valid and well known concern in HST parallels 
(\cf \cite{cas95}), and the obvious remedy is to excise these excess bright 
galaxies from the sample. These three galaxies were
therefore omitted from the final counts because they are not representative of
{\it field} galaxies in general, given that the primary HST target was biased
towards these clusters.

Also the brightest galaxy in the survey ($\Bj=16.99$
mag) was excluded from the counts because it is part of the NGC 383 group,
which was also the primary target of that particular HST acquisition. This
galaxy is an elliptical that appears to be interacting with a fainter spiral
galaxy, and it was only noticed because the ANN assigned a much later type. A
careful inspection shows that this galaxy contains a significant disk structure
with low level spiral arms which explains why it was incorrectly classified.
Therefore, there was only one true ``field'' galaxy found in this survey 
brighter than $\Bj\lesssim19$ mag. 

To let the readers judge our line of reasoning for themselves, we 
plot the three bright data points based on these four galaxies in the total
counts (upper left panel in Fig. 3), but plot their detected surface density as
{\it upper limits}, since the true field galaxy count at these flux levels
($16.5\lesssim\Bj\lesssim 18.0$ mag) must be lower than the count that we 
observe in these few not-so-random parallel fields. This is important for 
the discussion of the counts in \S ~\ref{results} and ~\ref{discuss}. 

\subsubsection{Misclassification Trends due to the Uncertain Rest-frame UV}
\label{uvmorph}

The mid-UV (2500-2900$\AA$) images of nearby galaxies available thus
far (\cite{win02}) suggest that it is more likely to misclassify true
ellipticals or early-type spirals in the rest-frame mid-UV as late type
galaxies than the other way around. This is because {\it truly} late-type
galaxies are dominated by young and hot stars in filters from the mid-UV
to the red, and so have to first order the same morphology and a rather small
morphological k-correction. However, early-type galaxies (ellipticals and
early-type spirals) can, {\it although do not have to} look dramatically
different when one goes from the rest-frame mid-UV to the optical-red part of
the spectrum. Hence, misclassifications due to the morphological k-correction
will more likely move more ellipticals and early-type spirals into the late
type/irregular category, rather than vice versa. Since all galaxies
in the present study were classified from $I$-band images, this would only
be relevant for galaxies with redshifts of $z \gtrsim 1.8$, which corresponds
to the fainter part of our sample, as can be seen in the morphologically
segregated redshift distributions of Driver \etal (1998). These redshift 
distributions show that significant numbers of $z \gtrsim 1.8$ galaxies are
only present in fainter sample where the classifications become increasingly 
unreliable ($\Bj\gtrsim27.0$ mag) for reasons discussed in \S~\ref{galclass}.
While this can perhaps explain part of the excess late-type/irregular galaxies
at faint magnitudes as discussed below, it cannot explain all of the FBG 
excess, and it certainly cannot explain the apparent excess of
early-mid types at $\Bj\gtrsim 24$ mag relative to the best available models,
as discussed below. 

\section{Results} \label{results}

Here we present our results for the $\sim1800$ BBPS galaxies as a function
of {\it measured} $B$-band brightness {\it and} ANN morphological type.
The statistically complete sample is composed of $\sim370$ galaxies with
$19\lesssim b_J \lesssim23.5$ mag.

\subsection{The Morphological Galaxy Counts} \label{morphcts}
\begin{figure}
\epsfig{file=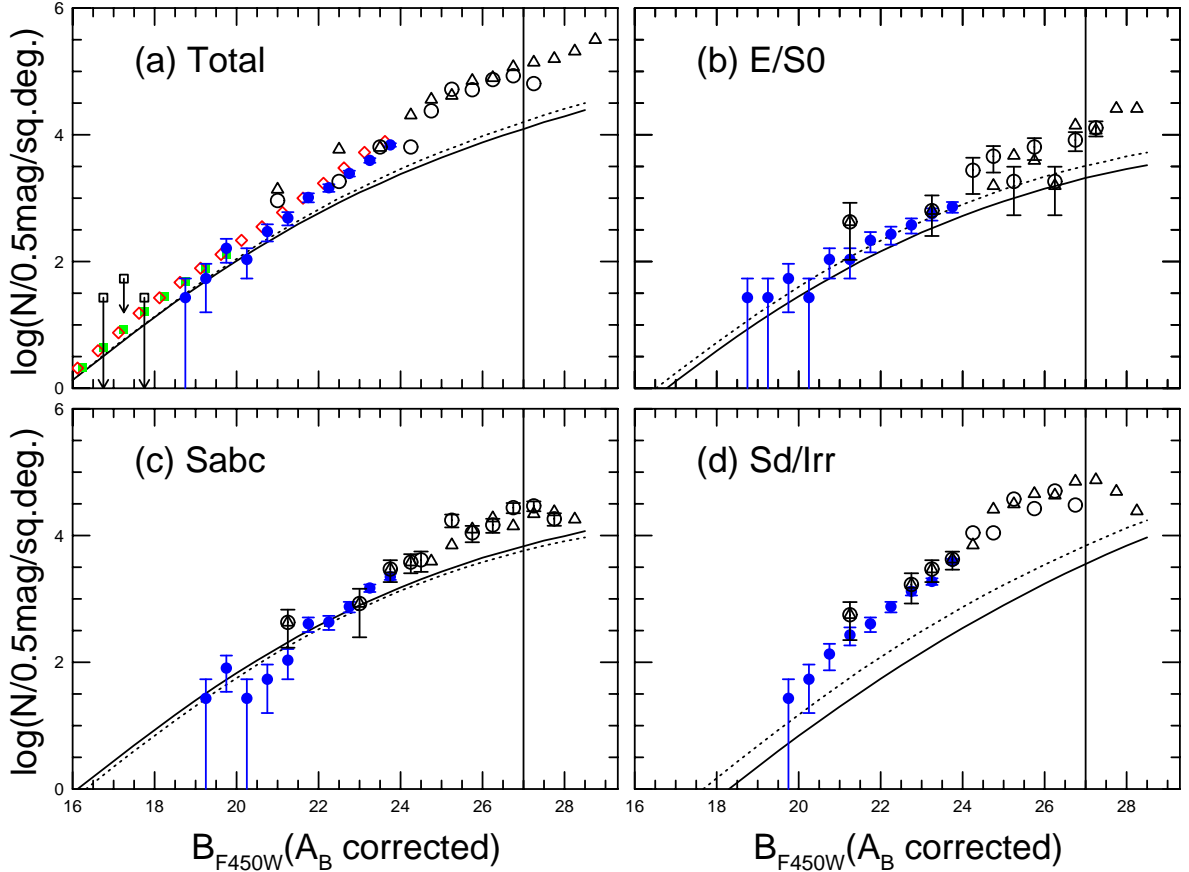,width=7.5in}
\figcaption[SHCohen.fig3.ps]{Differential $B$--band galaxy counts for all
BBPS galaxies as a function of morphological type and $b_J$ magnitude. The
solid blue points represent the new BBPS data that cover 
$18 \lesssim b_J \lesssim 24$ mag.  The open triangles (HDF--N) and open
circles (53W002) represent the counts from the two deeper fields analyzed
in the same way as the BBPS counts. The curves are predictions from local
LF plus no-evolution models as described in Odewahn \etal (1996) and in
Driver \etal (1995a). The models assume a cosmology with
($\Omega_{M},\Omega_{\Lambda}$) equal to ($0.3,0.7$). These non-evolving models 
use the
type-dependent LF's of Marzke \etal (1994; dashed lines) or Marzke \etal (1998;
solid lines). The open squares show the upper limits at the bright-end, as 
described in the text. The red diamonds are the total ground-based CCD counts
from the Millenium Galaxy Catalog (\cite{mgc02}) and the green squares are from 
a first part of the SDSS (\cite{yas01}). These SDSS and MGC total counts are
perfectly consistent. The reliable classification limit 
in the BBDS is $b_J\lesssim27$ mag, based upon the available training-sets and 
S/N in the images (see text and \cite{o96}). 
\label{galcounts}}
\end{figure}

\placefigure{galcounts}

The fundamental question we will address in this section is that of the galaxy
counts as a function of morphological type at fluxes between the
ultra-deep HST fields and what can be done from the ground. We will show
that, in order to use galaxy counts as a function of morphological type
to address issues of galaxy evolution, this intermediate flux range is crucial. 

In Fig.~\ref{galcounts}, we show the
BBPS galaxy counts as a function of morphological type. Also included for
comparison are the deeper counts from the two BBDS fields (\cite{o96}). At the bright end
($b_J\lesssim18$ mag), we include as upper limits the four galaxies,
discussed in \S~\ref{LLSS}, that were excluded from the sample due to their
high probability of not being representative of field galaxies. Also plotted 
are the combined ground-based counts from the Millenium Galaxy Catalog 
(MGC; \cite{mgc02}), which is a wide-field $B$-band CCD survey covering 30 
square degrees, which we have analyzed using the same software allowing for
the best direct comparison of our HST work to brighter ground-based data. 
For independent comparison, we also plot the transformed to the $b_J$-band 
counts from the SDSS Commissioning Data (\cite{yas01}). Both sets of 
ground-based counts match up well with the HST counts. The BBPS number count
data are tabulated in Table~\ref{tbl2}.

\placetable{tbl2}

The galaxy counts for the three main morphological types in 
Fig.~\ref{galcounts} now span a range of nearly 10 magnitudes 
($18.5\lesssim b_J \lesssim 28.5$ mag), of which about 8 magnitudes 
($19\lesssim b_J \lesssim 27$ mag) have decent statistics {\it and} reliable
classifications. This is a major improvement in survey dynamic-range that
could not have been achieved from the ground, since reliable classifications
are not possible for most galaxies in ground-based seeing for $b_J\gtrsim19$
mag (see \S~\ref{magre}). This also could not have been 
achieved from a few single deep HST fields, since these do not have 
sufficient statistics for $b_J \lesssim24$ mag. Thus,
the current combined HST morphological counts have the potential to set
significant constraints on galaxy evolution models in the flux regime
$19\lesssim b_J \lesssim 27$ mag. This is especially important 
for $b_J\gtrsim25.5$ mag, where routine spectroscopic measurements with 8--10 
meter class telescopes becomes increasingly difficult and incomplete.

The {\it total} counts are remarkably continuous and smooth from $b_J\sim18$ mag
down to the formal HDF detection limit of $b_J\sim29$ mag. In the flux range 
where both surveys have good statistics ($22\lesssim b_J \lesssim24$ mag),
the type-dependent counts for the BBPS and BBDS samples show good continuity, 
adding confidence that the $I$--band classifications of the faint BBDS 
galaxies are consistent with the brighter counts. 

For comparison with this new data, we plot in Fig.~\ref{galcounts} models
for the galaxy count models that are based on the local LF as a function 
of galaxy type, as described in Driver \etal (1995a), Driver, Windhorst 
\& Griffiths (1995) and Driver \etal (1998). These model have been updated
using the currently favored flat cosmology of 
$(\Omega_{M},\Omega_{\Lambda})=(0.3,0.7)$ and more recent estimates of the
local LF.  The solid lines use the type-dependent LF's of 
Marzke \etal (1994; hereafter CfA) and the dashed line uses the more recent
ones of Marzke \etal (1998; hereafter SSRS2). These are simple zero evolution 
models, which include only k-corrections with no explicit evolution in
luminosity, number density or color. The models in the upper left panel of
Fig.~\ref{galcounts} are the sums of those of the individual types in the other
panels. {\it No} LF-normalization has been applied at the bright end.
In fact, these new data show that such normalization is not needed, and that 
a global (\ie the same factor for all types) LF-normalization would cause the 
models to overpredict the observed counts for both the E/S0's and the Sabc's. 
Such an LF-normalization has been used in the past to help explain the excess of
fainter galaxies (for a discussion, see \cite{dwg95}; \cite{marz98}), but this 
is incompatible with the new data presented here. 

Therefore, the current data argue against a blanket LF-normalization with
the same amplitude for all types. This directly results from the fact that 
the new BBPS data yield both morphology {\it and} $b_J$-band magnitudes nearer 
to the normalization point ($b_J=18-20$ mag) than previous studies did.
The adopted cosmology, which allows for more volume at high redshifts than 
an Einstein--de Sitter ($\Omega=1$) model, has little effect near the
normalization point, but serves to bring the models {\it slightly} closer 
to the data at fainter fluxes (\eg the two models differ by a
factor of $\approx2$ at $b_J\simeq26$ mag).  From the high dynamic 
range of the current survey it is now becoming clear that the FBG
{\it excess} is almost entirely due to the late-type galaxies, which have the
steepest slope at the bright end, and are the most numerous type at the faint
end of the counts. In fact, we will argue that more than 90\% of the FBG excess
for $b_J \gtrsim 25$ mag is attributed to these late-types.
The next question to address is how to explain this. 

\subsubsection{Evolution Models vs. Renormalization}

As Fig.~\ref{galcounts} shows, the local LF's with zero-evolution do not 
reproduce the observed galaxy counts as a function of type. Though this
is most pronounced at the {\it faint} end, it is also true at the bright
end, especially for the late-types. This difference between the local
(\ie where the LF's have been measured) and the more distant Universe needs
to be further investigated. One way to model this discrepancy would be through
a normalization of the models, which is the equivalent of compensating for 
the possibility that the local present day Universe is either over- or 
under- dense in a particular type (or every type) of galaxy. Another way
is to assume some type of evolution (\eg luminosity or number-density) in
the galaxy populations at relatively low redshifts.

Under the assumption that the data do not justify an LF-normalization of the
models, especially for the early-types, we include luminosity evolution 
as a first step to explain the galaxy excess at faint magnitudes. As described 
in Driver \etal (1995a), we assume that galaxy luminosities evolve as
$L\propto(1+z)^{\beta}$, which is supported through the CFRS for early--mid
type galaxies (\cite{cfrs95b}). As local references, we use the LF's of the
SSRS2 (\cite{marz98}), which are the best available until the morphologically
type-dependent LF's from the 2dFGRS and SDSS become available. 
Fig.~\ref{galcounts} shows that using the CfA LF's leads to the same
qualitative conclusions in the discussion that follows. The new evolutionary 
models are plotted in Fig.~\ref{evolcts}. The $\beta=0$ (no evolution) models
are shown in black, $\beta=+1,+2,+3,etc$ (positive evolution) are shown in
green and $\beta=-1,-2$ (negative evolution) are shown in red.

The no-evolution models ($\beta\simeq0$) clearly provide reasonable fits
to the BBPS data for both the E/S0 and Sabc for $b_J\lesssim24$ mag. This
implies that the majority of these types with $b_J\lesssim23$ mag were in
place by the redshift $z\sim 0.5$, which is the median redshift at this
magnitude. There is an indication that models with a little more evolution
may better describe the data, but a small type-dependent LF-normalization
would also fit the data. From the bright-end to the limit of 
$b_J\lesssim23-24$ mag, ellipticals are well fit by $\beta\simeq1$
or a modest upward LF-normalization of +0.1 dex. Similarly, mid-type spirals
are well fit by $\beta\simeq$1--2 or an upward LF-normalization of
$\sim$0.1--0.2 dex for $b_J\lesssim24$ mag. The CFRS has shown that giant
early-mid--type galaxies underwent luminosity evolution with
$\beta \simeq$1 since $z\lesssim1$ (\cite{cfrs95b}), which further supports
our suggestion that a significant upward LF-normalization of the early--mid
types is not warranted.

As has been noted in the past (\cite{d95a}), the late-types are clearly 
in excess of the no-evolution models over the full flux range shown here.
In fact, even the most rapidly 
evolving model ($\beta=+5$), under-predicts the faint Sd/Irr counts at 
$b_J\gtrsim24$ mag. The best model for the total galaxy counts is the 
sum of this rapid evolution model for the late-types plus the no-evolution 
models for the earlier types ($\beta=0$), and is shown in green
in the upper left panel of Fig.~\ref{evolcts}. This model follows the 
observed total counts to the flux level $b_J\lesssim22-23$ mag, fainter than 
which the late-type irregulars cause the biggest discrepancy.
If one wanted to completely fit the {\it total} galaxy counts with a
no-evolution ($\beta=0$) model, one would have to normalize the 
late-type LF by a factor of $\approx4-5$, and such a large LF-normalization 
factor would be difficult to explain in any case as a deficit of galaxies 
in local surveys or inhomogeneties in local large-scale structure. A
combination of luminosity evolution and LF-normalization is also a 
possibilty to fit the late-type counts over the full flux range presented
here, but better statistics for $b_J \lesssim 20$ mag are required for
a stronger statement to be made in this regard.

In order to illustrate that the FBG {\it excess} is dominated by late-types, we 
consider two extreme cases. First, if one assumes that the zero-evolution, 
zero-renormalization models are correct, then the excess of the data
over the models for $25\lesssim b_J \lesssim 26$ mag is composed of
approximately 95\% Sd/Irr's. In the other extreme, where the E/S0's and
Sabc's are modeled by zero-evolution, zero-renormalization and the late-types
are renormalized a factor of 10 (an upper limit for illustrative purposes),
then the FBG excess in the same magnitude range is only slightly reduced to
93\% late-types. In other words, assuming that the models and classifications
are reasonable, both the faint counts and the FBG excess over the model
predictions are dominated by late-types.

One caveat that should be mentioned is that the morphological LF's from
the SSRS2 are separated by type (E/S0, Sabcd, Irr/Pec) differently than our
data where we place the Sd's in the late-type category. Assuming an 
upper-limit of Sd galaxies comprising 25\% of all spirals, this would
argue for lowering the models for mid-types by this much and raising the
late-type models by a similar amount. This is too small an effect to be seen
in the current data, but ultimately it is an issue that should be resolved.
Another possible explanation for the excess of late-types is that they were 
somehow excluded from the local surveys, many of which were based on
photographic data which have inherent low surface-brightness selection
issues. In fact, if one examines the data for the late-type LF
(\ie \cite{marz98}), it is seen that there are far fewer of this type of
galaxy, relative to the other types, in each luminosity bin. This means that
whether the lower SB galaxies exist and were undetected or don't exist at
all, the late-type LF is necessarily the least well determined of the three.
This potentially could be resolved with a new determination of the local LF's
from more modern data such as SDSS and 2dFGRS, although careful attention
must be paid to not select against these late-type galaxies which is clearly
a difficult problem to avoid. 

It is also readily apparent from Fig.~\ref{evolcts} that none of the
models fit beyond $b_J\gtrsim24$ mag. This probably implies that we are
seeing the epochs where galaxy evolution is more dramatic and not well
described by our model. Merging and hence number density evolution probably
play a major role in this regime and our models are therefore most likely 
too simplistic. The issue of merging and the effects of the $\Lambda$ are
further discussed in \S~\ref{speculation}.

The new BBPS data shown here make it very clear that the issues of
renormalization versus evolution cannot be disentangled without a
statistical sample of galaxies with morphological types that extend to
even brighter magnitudes. The steep slope of the late-type counts at the
brighter end ($b_J\lesssim20$ mag) indicates that filling this portion of
parameter space should provide a large step forward in modelling the
galaxy counts. Once this is done, different evolutionary scenarios
can be modelled and tested. Also, questions of merging and/or
morphological evolution should be further investigated, but this
requires the redshifts to be known (\cf \cite{cfrs4}).  Unfortunately,
the bright-end of the counts cannot be filled in by HST, because the
surface density of bright galaxies is too low to efficiently use the
small HST field-of-view. We will show in \S~\ref{summary} that ground
based efforts it $\Bj=19-20$ mag are limited by atmospheric seeing,
but is still plausible with the large CCD surveys that are now being conducted,
as long as the best seeing ($FWHM\lesssim1\farcs0$) images are used.

\subsection{The Magnitude--Effective Radius Relation} \label{magre}

\placefigure{bmagre}

The apparent effective radii of galaxies in the BBPS are shown in
Fig.~\ref{bmagre} as a function of $B$-band brightness and ANN type. Plotted is
the half-light or effective radius ($r_e$), averaged over the two pass-bands,
versus $B$-magnitude. This is the radius of the galaxy containing half the
light, with no assumption about the shape or type of profile. It is determined
from integration of the light-profile. The dark short-dashed lines represent
the approximate detection limits for the exposure depths in the two surveys. 
The detection limit for smaller, unresolved objects was modeled as a simple
Gaussian source convolved with the SExtractor convolution mask. The limit for
the more extended objects was modeled by an exponential disk. The fact that
the object detection is based on surface-brightness (\ie a minimum number of
pixels above a specified threshold) makes it readily apparent that bright
extended objects could be
missed by automatic detection procedures. This is further discussed below. The
colored lines that are almost vertical at the faint end give the expectations
for a local galaxy of a given absolute magnitude, morphological type and
intrinsic effective radius (as determined by a type-dependent relation between
absolute magnitude and size from the RC3) that is redshifted back to higher z.
These are the same models used in Odewahn \etal (1996), except that here we
used the $\Omega_M=0.3,\Omega_{\Lambda}=0.7$ and $H_o=65$ km/s/Mpc cosmology.

\placefigure{hist15}

The general trend in Fig.~\ref{bmagre} is that the brighter objects appear 
larger, even given the plotted completeness limits. More interestingly is 
that there are
separate, although overlapping, size distributions for the different
morphological types. This is shown in more detail in the histograms of
Fig.~\ref{hist15}. It is important to emphasize that size
information was not {\it directly} used in classifying the galaxies. As seen
in these histograms, the general trend is that, at a given flux level, E/S0
galaxies are the smallest in size while the Sd/Irr class contains the larger
galaxies. The arrows represent the observed medians of the distributions for
the BBPS (thin arrow, solid line) and for the BBDS (thick arrows, dashed line).
In rows where both data sets are shown, the BBDS histograms have been multiplied
by the ratio of the areas of the two data sets, while the actual BBDS numbers
are shown on the right axes. In all cases, the BBDS medians are (somewhat)
larger than the BBPS medians. The fact that the BBDS and BBPS medians do not
match in the second and third rows is due to the fact that the detection limit
results from the different surface-brightness completeness thresholds in
the two types of surveys, which reach rather different depths. The BBPS limit 
shown in Fig.~\ref{bmagre} shows that a galaxy with $b_J\approx24.5$ mag and
$r_e\gtrsim0\farcs8$ would escape detection in the BBPS, but would easily be
seen in the BBDS catalogs. Therefore, in rows 2 and 3 of Fig.~\ref{hist15}, the
BBPS median effective radii are {\it lower} limits, while the BBDS medians are
more realistic, but suffer from small number statistics. The bottom two rows of
the histograms come from the BBDS alone and suffer from similar SB-selection
problems as the BBPS has at brighter levels. In general, the main result is
that at a given brightness level, the
early-type galaxies appear smaller than the mid-type galaxies which appear
smaller than the irregulars, and that fainter galaxies in general appear
smaller than brighter ones. All this can be qualitatively seen in Fig. 5 ---
bearing in mind the respective SB-completeness limits of the different samples
--- and quantitatively in Fig. 6. 

The trend for the two samples is the same to the completeness limit of the BBPS
sample ($b_J\lesssim23.5$ mag). The two BBDS fields are also included because
they go deeper (formally $b_J\lesssim27.5$ mag for 53W002 and $b_J\lesssim29.0$ 
mag for HDF--N, see \cite{o96}), and because there are enough galaxies
past $b_J\gtrsim24.5$ mag to make the comparison. The reason that the BBDS
samples in Figs. 5--7 do not appear to go as deep in as they did in
Odewahn \etal (1996) is that these deep survey fields were re-analyzed 
in an identical way to the BBPS, in order to make the optimal comparisons
in the flux range where both surveys overlap. This includes the requirement
that, because of the limited number of parallel orbits available, an object
had to be detected in {\it both} the $B$ and the $I$-band filters in order
to be declared real. The middle panels ($23\lesssim\Bj\lesssim 25.2$
mag) for the three types clearly show that the larger BBPS objects are missing
at this flux level, which is due to the stated surface brightness detection
limits. The fact that the medians in the third row for the BBDS are larger 
than for the BBPS in the second row clearly illustrates that the completeness
limit is a function of size as well as brightness, i.e. a surface brightness
limit.

This surface brightness limit is very important because galaxy morphology is
correlated with surface brightness. The later-types are more extended than the
early-types at a given magnitude, and therefore are of lower surface
brightness. This would imply that the galaxy counts from the BBPS could be
underestimated for the later-types for $b_J\gtrsim 23$ mag. It is not clear
whether this is actually the case in the galaxy counts of Fig.~\ref{galcounts}. 
Although the BBPS counts are slightly lower at this flux level than in the 
deeper fields, which also have a better SB threshold, the difference is not
much larger than what we have seen from field-to-field variations. A careful
examination of Fig.~\ref{bmagre} shows that a small, but not insignificant
number of $b_J \lesssim 23.5$ mag lower surface-brightness (large $r_e$)
galaxies from the deeper fields would be missed in the BBPS due to the
different SB-thresholds {\it and} these are mostly later-typed galaxies.
Though only about a 10\% effect at $b_J \approx 23.5$ mag, it quickly
becomes a larger problem as one goes to fainter fluxes. Therefore, the
surface brightness selection criteria, which was implemented by specifying
a certain area above a threshold in SExtractor, could cause later-types to
be omitted, but that appears to be a minor effect for $b_J \lesssim 23.5$
mag in the BBPS.  However, this is an
important issue and the surface brightness--morphology correlation must be 
considered when detecting objects in this manner. Though presented here in the
context of our $b_J \gtrsim 19$ mag HST data, this issue is equally important
for both past and new local galaxy surveys, which need to be careful not to
select against late-type galaxies.
 
\subsection{The Color--Magnitude Diagram} \label{cmd}

\placefigure{cm}

Fig.~\ref{cm} shows the (B--I) color--magnitude diagram for this data. The two
BBDS fields were (re-)analyzed in exactly the same way as the 29 BBPS fields
to allow optimal comparison. It is important to re-emphasize that the
color was {\it not} an input parameter for the ANN classifier. Colors were
measured in matched and registered elliptical apertures whose sizes and shapes
were determined from the $B$-band images. For any given type, there is no tight
correlation of color and magnitude that might be expected if one were observing
a uniform population at different redshifts, suggesting that we are seeing
a scatter which is cosmic in origin.

In general, the $B$ vs. $(B-I)$ color--magnitude diagram shows that the
reddest galaxies are E/S0's while the bluest ones are Sd/Irr's, with the Sabc's
filling in the broad part middle of this range. However, there is also a
fraction of red galaxies that are classified as ``late-types.'' These could
either be misclassifications, or possibly galaxies with {\it some}
star-formation that is reddened by dust.

There are also some very blue galaxies that have been classified as
ellipticals, as also noted by Driver, Windhorst \& Griffiths (1995), who used
$(V-I)$ colors. This could result from several factors. First, they could be
misclassified objects. This is unlikely at brighter flux levels, unless the
profile is contaminated by a neighboring galaxy, because ellipticals are so
compact, which causes the S/N in the light profile to be rather good. 
Secondly, some of these, especially the brighter ones, may be misclassified
and/or possibly saturated stars. Visual inspection shows this to be
unmistakably true only for the few bluest outliers. The third possibility is
that some of the ``elliptical'' galaxies have an additional and significantly
younger stellar population. This could be indicative of a burst of recent
star-formation, as is seen in Compact Narrow Emission Line Galaxies (CNELGs,
\cf \cite{koo95}). A fourth possibility, is that these objects, especially 
those with a small measured $r_e$, could have a significant AGN component.
Lastly, this could be similar to the recent evidence against the traditional
single burst model for star-formation in elliptical galaxies
(cf. \cite{glaz98}, \cite{jim99}). This would require significant ongoing
star-formation (possibly induced by a recent minor merger), while the
light-profile of the product had already sufficiently settled into an
$r^{1/4}$-like profile. 

To test these conjectures we have adapted the models of Windhorst \etal (1994),
which were based on those of Windhorst \etal (1985) and 
and Kron, Koo \& Windhorst (1985), for our filter set and cosmology. In each
panel of Fig.~\ref{cm}, we have plotted these models for $M_B=-20.7$ mag which
is approximately $L^{*}$ for this cosmology. The solid lines are for ages
(\ie last major epoch of star formation) of 14 Gyr, the dashed lines are for 13
Gyr and the dot-dashed line is a non-evolving model. The red lines are 1-Gyr 
early-burst or $C$-models that would represent early forming E/S0's that form
all of their stars during the first Gyr after formation begins. The green lines
are passively evolving $\mu=0.7$ models, where 70\% of the mass forms into 
stars in the first Gyr with an exponentially declining SFR and is 
representative for the mid-type Sabc's. The blue
lines are actively evolving $\mu=0.4$ models, where only 40\% of the mass
forms into stars in the first Gyr, a reasonable approximation for Sd/Irr's. 
We plot all models for each morphological type to allow a comparison of the
morphologies with star formation histories.

Note that some galaxies appear ``redder'' than the red/green upper envelope
model. Another way to look at this is that they are instead more luminous
than the $M_B=-20.7$ mag models plotted. Had one plotted instead the brighter
$M_B=-21.7$ mag models, or an even brighter model like $M_B=-22.2$ mag, these
objects would also have been represented by these upper-envelope models (\ie the
models would simply shift upward/brighter by $-1.0$ or $-1.5$ mag). The same
is true for the blue model representing the late-type objects --- they do in
general explain the observed blue objects rather well except for the few very
blue objects seen at $(B-I) \lesssim 0.0$ mag. Those would generally require
later star formation than is present in the $\mu=0.4$ model, have a weak 
AGN component and/or possibly have larger photometric errors. 

Therefore the observed spread in color present in the data is roughly
represented by a reasonable range of plotted models (the $C$-models though
the $\mu=0.4$ models), except for the very reddest and bluest objects. The 
former can be explained by assuming $M_B=-21.7$ mag to $M_B=-22.2$ mag, the 
latter by assuming a more constant star formation scenario (perhaps 
$\mu=0.1-0.2$). 

\section{Discussion} \label{discuss}

Here we discuss the properties of the three morphological types as seen in the
present BBPS data and compare it to other studies. The discussion will focus
on the brighter BBPS galaxies, since that is the main subject of the present
paper. 

\subsection{E/S0's} \label{early}

Adding the brighter elliptical counts from the BBPS does not substantially 
change the conclusion of Odewahn \etal (1996) that these counts can be modeled
rather well by the no-evolution predictions for $b_J\lesssim 25$ mag. The new
data strengthens this conclusion and also argues against re-normalizing the
models of the E/S0 galaxy counts, although there is the hint that these
galaxies may be undergoing some evolution at the faint end
($b_J\gtrsim24$ mag). As discussed in \S~\ref{intro}, the issue of evolution
of the ellipticals, which in principle should be the simplest to study, is 
far from settled. It is clear that the high resolution of HST is needed to
select a sample of field ellipticals, and that redshifts for all objects are
also needed. This was done for 46 galaxies from the CFRS/LDSS (\cite{cfrs3}),
which showed that field ellipticals are not composed entirely of old stellar
populations, but were largely in place since $z\simeq 1$ and therefore support
the view of an early formation epoch with occasional, more recent episodes of
star-formation.

Though the reddest galaxies at a given magnitude are generally of early-type,
Fig.~\ref{cm} shows that there is a broad range of colors for galaxies of
elliptical morphologies. The same conclusion was reached by the CFRS/LDSS, who
computed rest-frame ($U-B$) from the observed ($V-I$) using the measured
redshifts. As previously mentioned, the bluer ellipticals could be similar to
the CNELGs (i.e. compact and blue), which are thought to be distant analogs of
local HII galaxies as well as progenitors of today's spheroidal galaxies
(\cite{koo95}).

The size distribution in Fig.~\ref{hist15} shows the well known fact
that the ellipticals
are the most compact galaxies of all Hubble types. The average effective radius
for $b_J\lesssim24$ mag is $r_e \simeq 0\farcs35$. It was also shown in
Fig.~\ref{hist15} that their effective radii increase with brightness.

Overall, these results are somewhat contrary to the view that {\it all}
ellipticals are just old fossil galaxies, which all formed at the same time,
whose stellar populations are simply
passively evolving or are well understood. While this seems true in the
local Universe, the evidence is mounting against old red ellipticals at
higher redshift, where there seems to be a larger color scatter for all types
than is seen locally at the same rest-frame wavelengths. Even in high redshift
cluster studies (e.g. at $z=0.83$; \cite{pgd00}), there are cluster galaxies
with redder colors that are not of elliptical morphology. Therefore, it is
dangerously uncertain to use a {\it single}
color for high redshift galaxies as an indicator of morphological type.
However, it may still be possible to determine types by using multiple colors
which means that the galaxy spectrum is being sampled in more than two places
across the SED, assuming that the spectra of the objects are well understood.
Given that there seems to be some discrepancy between the properties of the 
local and distant elliptical galaxies, a comprehensive comparison of their
SED's seems warranted in order to further investigate the situation.

\subsection{Sabc's}

The counts of the BBPS spirals also do not seem to require a large amount of
luminosity evolution nor a renormalization to fit the models based on local
LF's.  There are far more spirals than ellipticals, but the numbers of
mid- and late-types are about the same for a given magnitude. The sizes of
the spirals are larger than the early-types and they have a median of
$r_e \simeq 0 \farcs 41$ at $b_J\lesssim24$ mag.  It is important to
re-iterate that a comparison of the deeper BBDS sample and the BBPS in a
flux range where they overlap shows how the surface-brightness selection 
biases the sample {\it against} the larger galaxies at a given magnitude.
This is a result of defining the object detection limit as a minimum area
above some threshold SB. This is seen in Fig.~\ref{hist15}, where the
higher signal-to-noise BBDS fields have a larger fraction of larger mid- and
late-type galaxies than the BBPS for $b_J\gtrsim 23$ mag. Lilly \etal (1998)
showed that the distribution of the physical sizes of larger ($\alpha^{-1} >
3.2$ kpc) disk galaxies is roughly constant for $z\lesssim 1$. However, a study
of the HDF--N disk galaxy sizes showed an excess (in relation to a CDM model
prediction) of faint ($M_B >-19$ mag) disk galaxies with smaller sizes ($R_d <
2$ kpc) seen from $z\lesssim3$ even down to $z \sim 0.5$ (\cite{gia2000}). It is
clear that the issue of size evolution and the growth of disks in spirals is
far from settled, and the significant number of BBPS spirals found in the
current work will allow further spectroscopic studies to elucidate the physical
properties of these objects. 

The observed colors of the spirals show a similar distribution to the
irregulars. The spiral distribution is also broad, but the median is bluer than
that of the early-types --- $(B-I)_{med}=2.8$ mag for E/S0 versus
$(B-I)_{med}=2.1$ mag for Sabc for $b_J\lesssim 24$ mag. This large dispersion
in color is likely due to an intrinsic difference in the galaxies, and not
entirely to bandpass-shifting. This is apparent in the redshift dependence of
the ($B-I$) color given by Roche \etal (1997), which shows that the dispersion
in color expected for a given type is small for $z\lesssim 1$, especially for
the later-types. A large dispersion in the rest-frame $(U-V)$ colors was also
shown to exist in the disk dominated galaxies of the CFRS/LDSS with only a weak
dependence on redshift (\cite{cfrs2}).

\subsection{Sd/Irr's}

The counts of galaxies beyond $b_J\gtrsim24.5$ mag are
dominated by the later-types. An examination of the counts of the brighter
BBPS galaxies shows that somewhat brighter than $\Bj\simeq22$ mag, the counts
become dominated by spirals. It will be very important to carry out this study
to even brighter magnitudes ($\Bj\lesssim 19$ mag --- with HST or from the 
ground in
the best possible seeing) to see exactly where this crossover occurs. This
will be very near the region where the count models are normalized ($b_j
\approx 18$ mag), which makes it all the more important to understand the LF
normalization as a function of galaxy type.

The Sd/Irr's are generally more extended than the spirals and ellipticals and,
by definition, have a less regular appearance. At $b_J\lesssim24$ mag, their
median size is $r_e \simeq 0 \farcs 61$ and the SB selection effect mentioned
above is therefore even more important here. The color distribution is very
similar to the spirals. In fact, the bright spirals seem very similar to the
bright late-types in terms of number counts, colors and sizes. However, given
that they are classified differently by the ANN, they have differently shaped
azimuthallty averaged light profiles. These Sd/Irr light profiles are similar
to those of galaxies that have been visually classified as late-types
(i.e. the training set of \cite{o96}). The ANN types used here are based on a
system that does not have a merger class. Mergers are most effectively studied
by combining the high resolution imaging with measured redshifts (\cite{cfrs4})
to rule out chance superpositions of pairs. Hence, without redshift information
for all 1800 BBPS galaxies to $\Bj=24.5$ mag, we will defer studies of the
pair-fraction and merger rate until spectroscopic or good photometric
redshifts become available.

\subsection{A Possible Explanation for the Excess of All Types at
$b_J \gtrsim 24$ mag} \label{speculation}

\begin{figure}
\epsfig{file=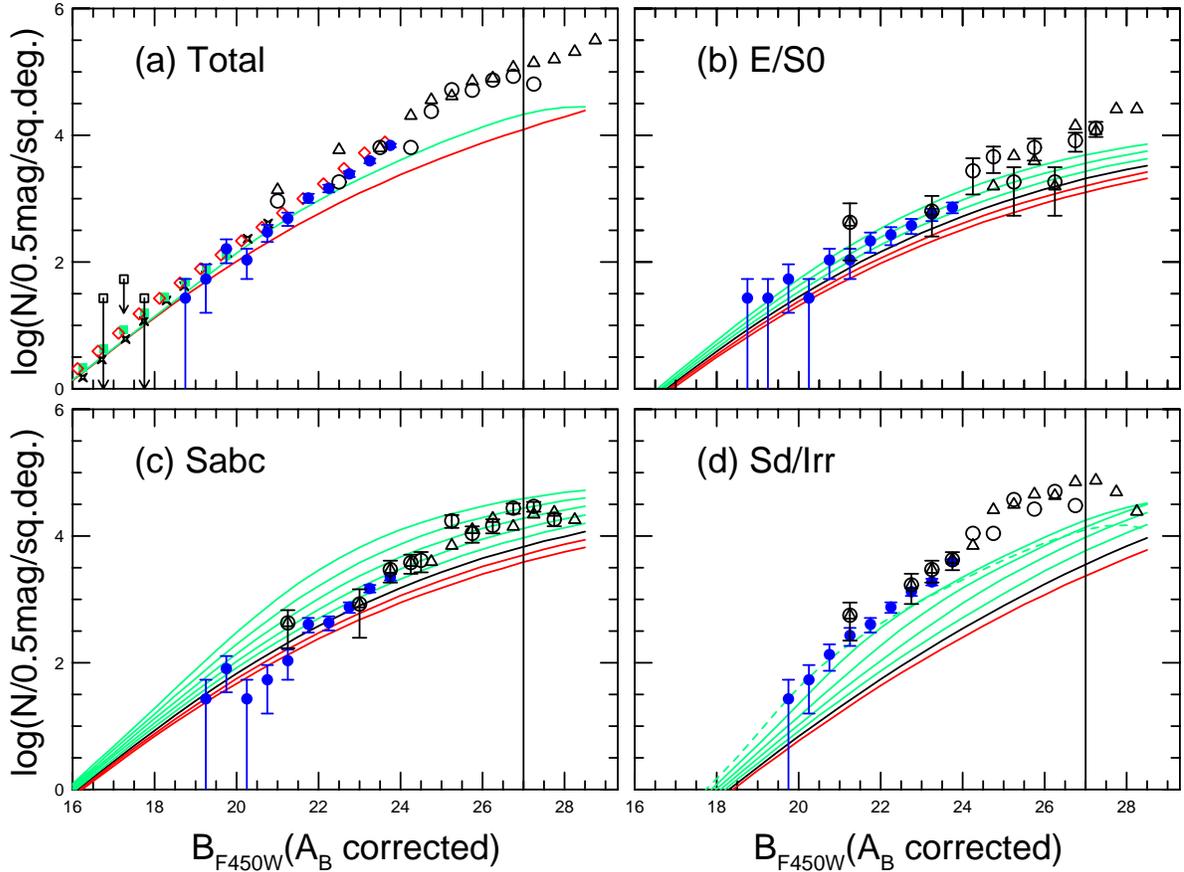,width=7.5in}
\figcaption[SHCohen.fig4.ps]{Evolutionary models for the differential 
$B$--band galaxy counts as a function of morphological type and $b_J$
magnitude. The solid blue points represent the new BBPS data that cover
$18 \lesssim b_J \lesssim 24$ mag.  All plotted data symbols are the same
as in Fig.~\ref{galcounts}. These models use the LF's of Marzke \etal (1998),
and assume galaxy evolution in the form of $L\propto (1+z)^\beta$. The
no-evolution ($\beta=0$) models are shown as solid black lines, positively
evolving models with $\beta=+1,+2,+3,+4,+5$ as green lines, and negatively
evolving model with $\beta=-1,-2$ as red lines. In panel (a), the red line
is for $\beta=0$ for all types, and the green line is for $\beta=0$ for
the E/S0 and Sabc galaxies and $\beta=+5$ for the Sd/Irr population.
Clearly the Sd/Irr counts in panel (d) are well above any reasonable
model for $b_J \gtrsim23$ mag, and this is causing most of the {\it excess}
of the total counts in panel (a).  \label{evolcts}}
\end{figure}

Perhaps our most curious result is that none of our new models fit the observed
galaxy counts as a function of type at flux levels fainter than
$b_J\gtrsim24$ mag (Fig.~\ref{evolcts}). This is despite the fact that
these models now contain the latest $\Lambda$-dominated cosmology and the
best available local LF as a function of type. For all galaxies, there
appears to be a significant excess for $24\lesssim b_J \lesssim 27$ mag,
which corresponds to a range in median redshift of approximately 
$\zmed \simeq0.5$ to $\zmed \simeq1-2$. This excess amounts to up to a
factor 3--4 for the early--mid types, and up to a factor of 6--10 
for the late types with
respect to the local LF plus the best fit luminosity evolution model for
$b_J \lesssim 24$ mag (\ie $\beta=1$ for early--mid types and $\beta=5$
for late types). This appears to be a robust result, since the model
galaxy counts as a function of type approximately fit the brightest
available data points, which are now around $b_J \simeq19-20$ mag.

In this section we suggest a possible explanation for this, namely that we are
witnessing a global and type-dependent excess of galaxies of {\it all} 
types at $b_J \gtrsim 24$
mag, where the median redshift is $\zmed \gtrsim 0.5$ --- with a larger
excess for the late types --- and explore possible physical causes for
this excess.

Together with the small galaxy sizes (Figs.~\ref{bmagre} --~\ref{hist15})
seen at faint magnitudes, this excess suggests that faint galaxies may be
more numerous and smaller at $z\simeq1-2$ compared to the ones seen today.
While the luminous disk galaxies seen at $z\lesssim 1$ have approximately
the
same size distribution as that seen today (\cite{cfrs2}), this may no
longer be true for $z \gtrsim 1$. Hierarchical formation scenarios
predict
larger numbers of smaller objects in the epoch $z\simeq1-2$  and beyond.
Based on the redshift distribution as a function of morphological type,
Driver \etal (1998) suggest that an excess of mid--late type galaxies 
is seen for 
$I \gtrsim 23$ mag ($b_J \gtrsim 24.5$ mag), especially in the redshift 
range $z\simeq1-2$. They
tentatively identified this epoch as the epoch of disk formation. If
true, then both the intrinsically smaller disk galaxies at $z\gtrsim 1$,
and
the temporary end-products of the mergers (\ie the early-types), as well
as the building blocks and mergers-in-progress with temporarily enhanced
star-formation (\ie the late types/irregulars/peculiars) may all be
enhanced in numbers past $z\gtrsim 1$, and also be smaller in size and
mass.

Galaxy formation may have proceeded such that, in the redshift range
$z\simeq1-2$ and at higher redshifts, {\it both} the early- {\it and}
mid- {\it and} late-types were present, but likely in proportions that
{\it slowly} changed with cosmic time, and likely with considerable
migration back and forth between galaxy classes. For example, mergers
between two spiral disks, or between a spiral and an irregular would
eventually result in a bulge dominated galaxy, but since star-formation
is known to not be a very efficient process, not all gas would be used up
during that merger, nor would  that gas necessarily reach escape velocity,
so the gas would eventually settle back as a bulge dominated galaxy with a
newly re-formed disk (\eg \cite{hib96}). Hence, the end-product of a merger
would temporarily be a bulge-dominated galaxy, but it could grow a disk back
in the next few gigayears after the merger, and then look like a spiral galaxy
until the next (major) merger occurred. Similarly, many luminous irregulars
and peculiars may be the temporary stages when observing mergers in
action before a system settles as a bulge-dominated or a disk-dominated
galaxy (\cite{bar96}), although a good fraction of the late types that we see
may likely just be the numerous smaller galaxy building blocks from which the
hierarchical merging started (\cite{p96}). In conclusion, in the
hierarchical scenario, there would be considerable migration back and
forth between galaxy classes, and galaxies at high redshifts would be
smaller and more numerous than those seen locally (Fig.~\ref{evolcts}),
consistent with the excess seen for {\it all} types at $b_J \gtrsim 24$
mag (Fig.~\ref{evolcts}b--~\ref{evolcts}d).

The one remaining issue that begs an explanation is what physical mechanism
could explain the larger numbers of {\it all} types at $b_J \gtrsim 24$ mag,
or $z \gtrsim$ 0.5--1.0? In hierarchical formation scenarios 
(\cite{nfw96}, etc), bulges form relatively quickly in the epoch $z\simeq3-5$
and mostly via major mergers. Disks form later in the epoch $z\simeq1-2$,
but more through the gradual (hierarchical) infall gas or minor
mergers. These predicted scenarios can be seen in the morphological
redshift distributions of Driver \etal (1998). The merger rate was higher in the
past by $(1+z)^m$ where $m\simeq$ 2--3 (\cf \cite{burk94}; \cite{neu97};
\cite{cfrs4}), but mostly so for $z \gtrsim$ 0.5--1.0. For the
currently accepted values of $\Omega_m\simeq0.3$ and
$\Lambda\simeq 0.7$, the $\Lambda$-driven acceleration starts dominating 
the expansion of the Universe for the first time at $z\lesssim 1$. We
hypothesize that --- as a consequence --- the galaxy merger rate
gradually winds down in the epoch $z\simeq0.5-1.0$. For instance, groups
of smaller galaxies or sub-galactic units that were nearly virialized at
$z\gtrsim 1$ will still virialize for $z\lesssim 1$, but groupings of such
objects that were not even close to turn-around at $z\simeq1$ will be still
expanding with the Hubble flow at $z\simeq0.5$, and probably forever do so
in a $\Lambda$-dominated universe. At $z\lesssim 0.5$, these late-types
would never do much further merging, but just fade away (\cf
\cite{fb98}). The end-result is that one observes a slowly evolving
Universe consisting of E/S0's and Sabc's for $z\lesssim 0.5$ (or
$b_J\lesssim 24$ mag) --- as we observe here in 
Fig.~\ref{evolcts}b --~\ref{evolcts}c --- plus the
relatively rapid dwindling-away of late-types for $z\lesssim$ 0.4--0.5,
explaining their steep counts for $b_J\lesssim 24$ mag (Fig.~\ref{evolcts}d).
And one would observe a {\it vast} increase in numbers for all types at
$z\gtrsim 0.5$ (or $b_J\gtrsim 24$ mag), and especially at $z\gtrsim 1$
(or $b_J\gtrsim 26$ mag), because most groupings of smaller objects had
still plenty time to turn-around from the Hubble flow and overcome the
effects of $\Lambda$ at $z\gtrsim 1$. Since the merger rate was likely
much higher at $z\gtrsim$ 0.5--1.0 than at $z\sim$ 0--0.4 (\cite{cfrs4}),
merging proceeded rapidly and successfully for
$z\gtrsim 0.5$, and vastly reduced the galaxy numbers with time at
$z\simeq$ 0.5--1.0, so that larger numbers of {\it all} types are seen
at $z\gtrsim 0.5-1.0$, with the largest increase for the late types.

In conclusion, the new BBPS data shown here makes it clear that the issues of
LF-normalization versus evolution cannot be disentangled without a
statistical sample of galaxies with morphological types that extend to even
brighter magnitudes. The steep slope of the late-type counts at the brighter
end ($b_J\lesssim20$ mag) indicates that filling this portion of parameter space
should provide a large step forward in modeling the galaxy counts. Once this
is done, different evolutionary scenarios can be modeled and tested. Also,
questions of merging and/or morphological evolution should be further 
investigated with better statistics from larger surveys, with wider dynamic
range, and through a systematic assessment of the effects from the uncertain
rest-frame UV (\cite{win02}) on the classifications at the {\it faintest 
magnitudes}.

\section{Summary and Future Work} \label{summary}

\begin{figure}
\plotone{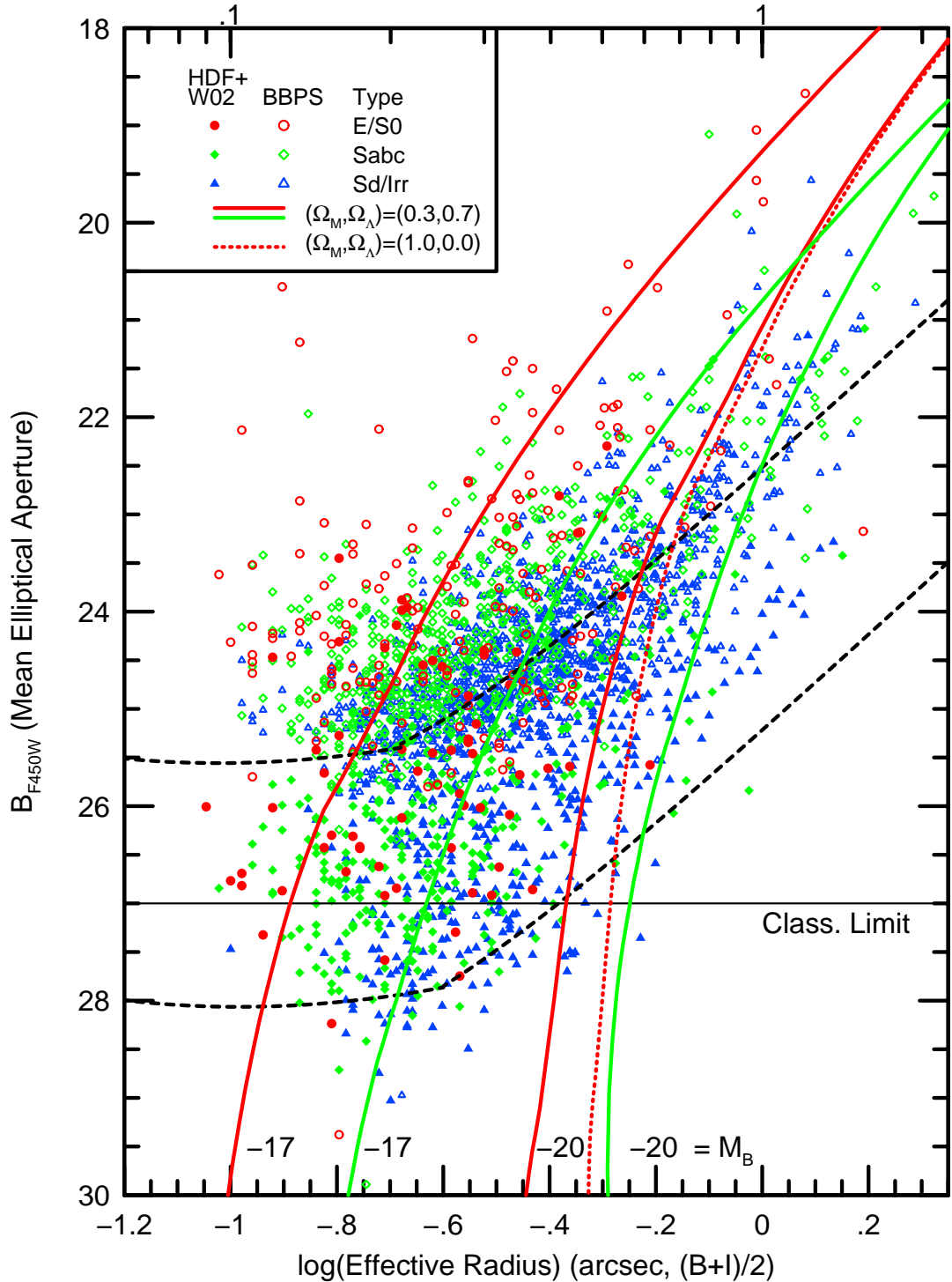}
\figcaption[SHCohen.fig5.ps]{$B$--band magnitude--effective radius relation for
galaxies in BBPS and BBDS data sets. The thick dashed lines represent the
approximate completeness limits for the deep and shallower surveys as described
in the text. The lines that are almost vertical at the faint end from Odewahn 
\etal (1996) show the expected $\Bj$--\re relation for redshifted RC3 galaxies
of a given absolute magnitude and assumed cosmology, as indicated. Objects
below the horizontal line at $b_J=27$ mag are beyond the reliable classification
limit of the BBDS. The scale across the top indicates $r_e$ in arcseconds.
The quantity $r_e$ plotted here is the average of the individual effective
radii measured from the light profiles in the $B$ and $I$ pass-bands. The
$r_e$ values are in general very similar between the $B$ and $I$ filters, so
yielding a better measure of $r_e$. \label{bmagre}}
\end{figure}

\begin{figure}
\plotone{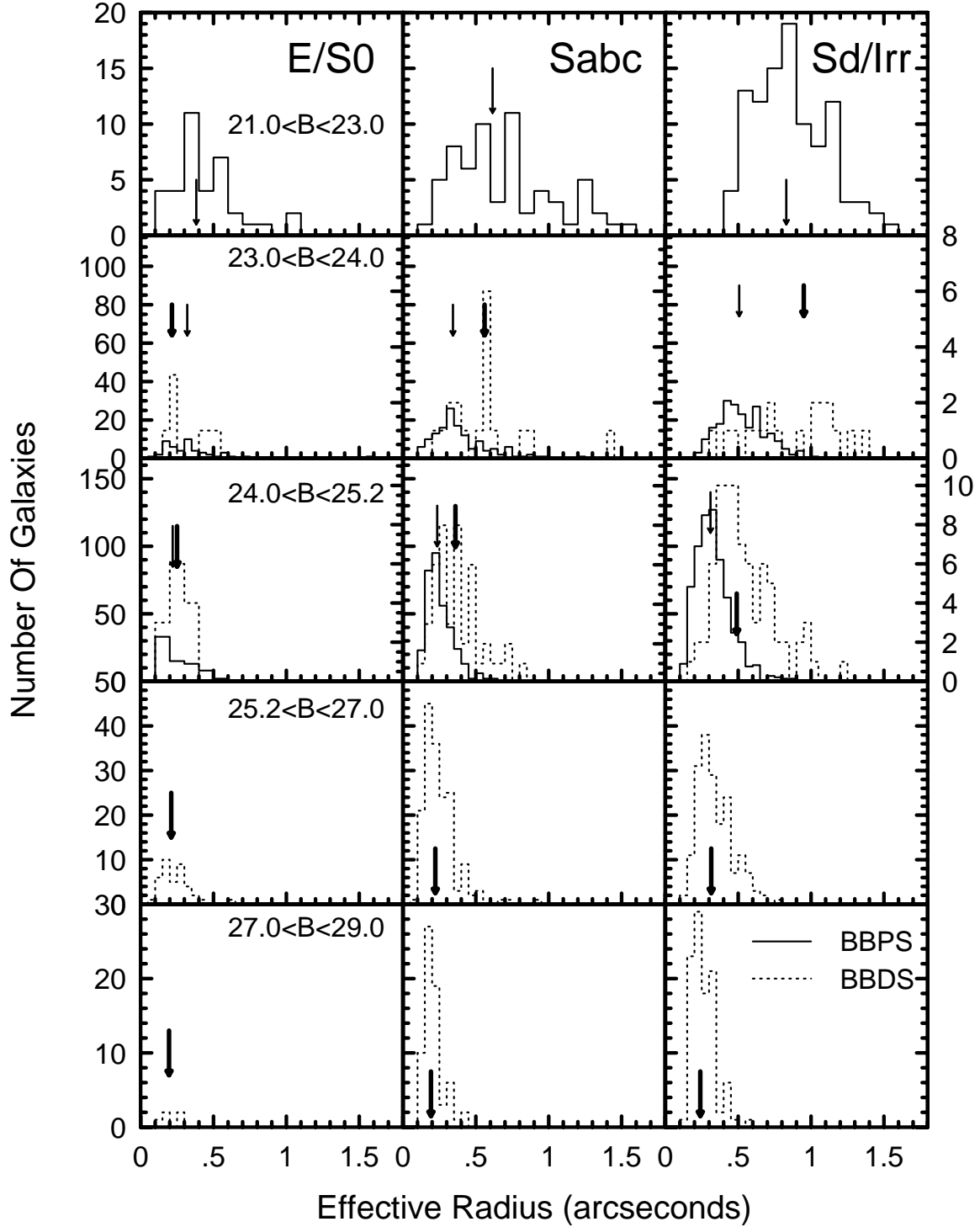}
\figcaption[SHCohen.fig6.ps]{Distributions of galaxy sizes as a function of
$B$--band brightness and $I$--band morphological type. The arrows indicate the 
median effective radii for the given distributions. The solid histograms and
thin arrows are for the BBPS (using the number scale plotted on the left), and
the dotted lines and thick arrows are the BBDS (using the scale on the right).
In the second and third rows ($23\lesssim\Bj\lesssim 25.2$ mag), the BBDS
data has been scaled up by the ratio of the areas of the two data-sets for
comparison purposes. The completeness of the histograms as a function of
flux and size is discussed in the text. \label{hist15}}
\end{figure}

\begin{figure}
\plotone{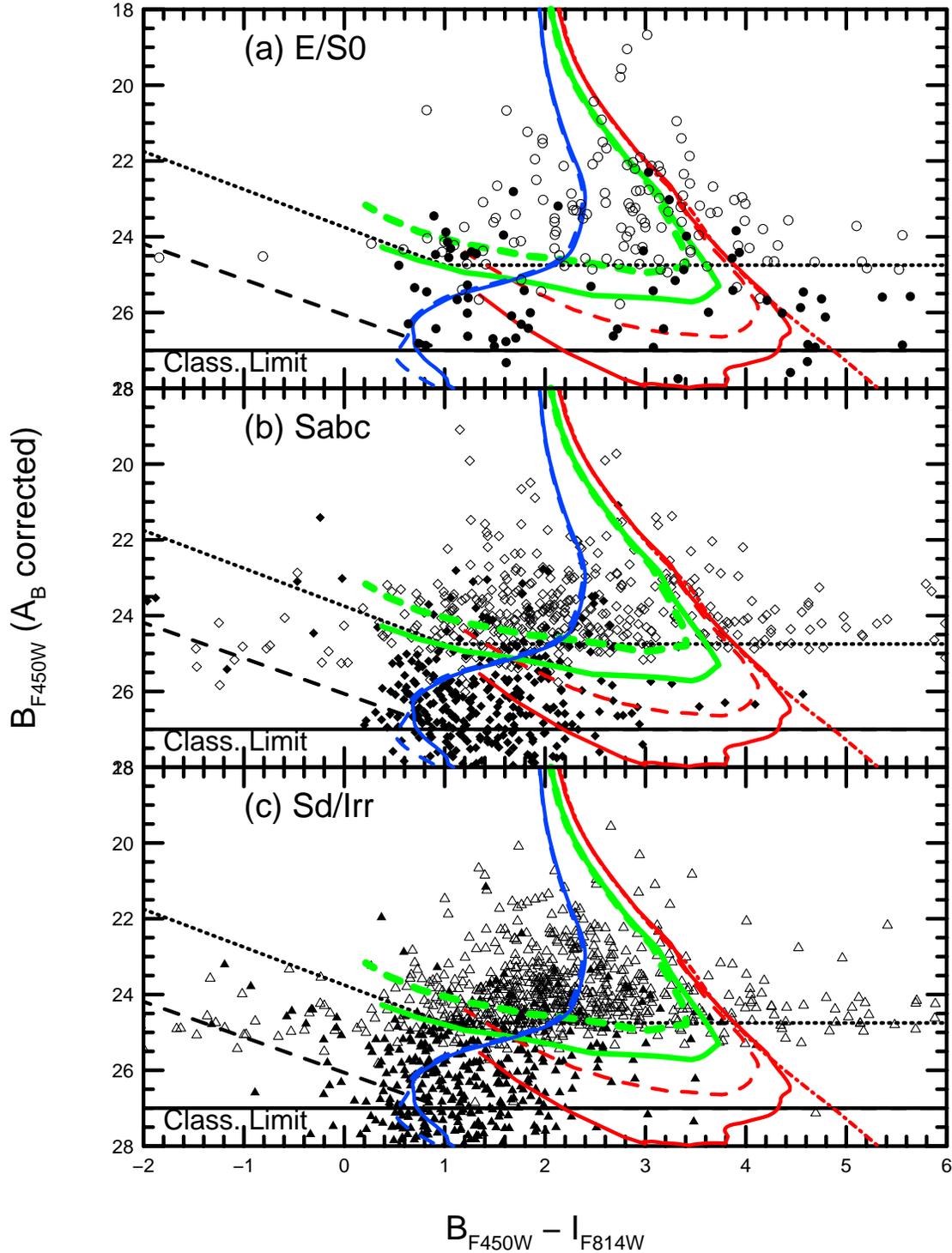}
\figcaption[SHCohen.fig7.ps]{The $B$ vs. $(B-I)$ color magnitude diagram as a
function of galaxy type. Each of the three different morphological ANN types
are plotted in different panels as indicated. Both the two BBDS (solid symbols)
and 31 BBPS (open symbols) fields were measured in exactly the same way for
consistency. The slanted and horizontal black lines indicate the $B$--band
and $I$--band 50\% detection limits, respectively. The solid black line at
$b_J=27$ mag indicates the $I$--band classification limit (as limited by the
uncertain rest-frame near-UV morphology and S/N considerations). The models
are plotted for a $M_B=-20.7$ mag galaxy that is either 14 Gyr old (solid) or
13 Gyr old (dashed). The non-evolving model is plotted as a red dot-dashed
line. The red, green and blue lines are meant to be representative of
E/S0, Sabc and Sd/Irr galaxies, respectively, although they are not
neccessarily unique. More details of the models can be found in the text.
\label{cm}} 
\end{figure}

We have presented an HST survey that connects the extremely deep HST studies,
such as the HDF--N ($24\lesssim\Bj\lesssim 29$ mag) and existing ground based 
studies such as the RC3 ($\Bj\lesssim 17$ mag). For example, the Stromlo-APM 
Redshift Survey
(\cite{lov96}) provided a catalog with morphological types for 1797 galaxies
and is complete to $b_j=17.15$ mag, although there is some question as to the
reliability of these catalogs (\cite{poz96}). Unfortunately, there is a
relatively small amount of good morphological data available in the literature
for a magnitude range of $17.15 \leq b_J \leq 19$ mag. The bright-end of the 
galaxy counts cannot be filled in by HST because the surface density of 
bright galaxies is too low to efficiently use the small HST field-of-view. Our 
expectation is that this flux range can be addressed from the ground with 
existing telescopes and larger area detectors in good seeing. The SDSS 
and other wide angle ground-based CCD surveys should soon provide this 
data. The key problem with these surveys will be in classifying the 
fainter ($\Bj\gtrsim 19$ mag) galaxies from the ground due to seeing-related 
effects, since their median \re values are $\lesssim 1\arcsec$ 
(see Fig.~\ref{bmagre} here), and rapidly decrease towards fainter
fluxes. Hence, the flux range $19\lesssim\Bj\lesssim 24$ mag must be studied 
with HST, which was the purpose of the current study. 

Another important piece of missing information is the redshifts of the BBPS.
Since the BBPS galaxies are brighter than $b_J\lesssim25$ mag, this is a project
that can be started on a 4 meter class telescope and finished on an 8--10 meter
class telescope. Objects of known redshift (even if estimated through
photometric redshifts) can, in principle, be more accurately classified,
because one can more
effectively correct for the effects of band-pass shifting, mentioned in
\S~\ref{galclass}. The BBPS along with measured redshifts can provide a wealth
of information when combined with the wealth of HDF--N redshifts which are now 
measured (\cite{jcoh00}).  Given the photometric redshift distributions of
Driver \etal (1998) for the HDF--N, it is clear that the majority of the
galaxies at $b_J \lesssim 25$ mag studied in this paper are
at $z\lesssim1$. Therefore, in order to classify galaxies observed in the
$B$--band, we need to know what local galaxies look like in the rest-frame
$U$--band or at slightly shorter wavelengths (2500--3000$\AA$). 
Relatively recent studies of the near--UV morphology of nearby galaxies show
some morphological differences as compared to the $B$-band (\cite{glb96};
\cite{bwo97}; \cite{kuch01}; \cite{marc01}; \cite{win02}), but a full
quantitative analysis of how this would affect the galaxy counts has yet
to be performed. The $I$-band classifications used in the present paper
largely avoid this issue, except for the relatively small number of higher
redshift ($z\gtrsim1.5$) galaxies whose classifications are therefore 
necessarily uncertain.

The use of artificial neural networks was previously shown to be effective in
classifying a large number of galaxies in a quantitative and systematic (\ie
reproducible) way (\cite{o96}). This method, based almost solely on the shape
of the measured light-profiles, has been applied to our much larger data set at
brighter levels. The appeal of this method is that it uses a large number of
photometric parameters, and that it is also ``trained'' based on human
classifiers in an effort to categorize the actual appearance in a systematic,
albeit non-linear way. Other automated techniques in use today can produce 
consistent results and provide other useful types of information, while also 
providing a good consistency check on the method used here. The next logical 
step is to see if the ANN method can be improved by using some of the 
2--dimensional information in the images. The ultimate goal, which no 
published method to date has achieved for faint galaxies (including the
one used in this paper), 
would be to quantitatively measure the true morphology in an automated way. 
This would involve distinguishing between more subtle features such as spiral 
arms, bars and rings, as well as differentiating between, for example, Sa 
and Sb, and tracing their behavior with redshift (see \cite{O02}).

In summary, the galaxy counts, size distribution, and $(B-I)$ color
distributions seen in the deep HST studies are consistent with what we are now
seeing with good statistics for the brighter BBPS galaxies. The {\it excess} of
faint galaxies for $b_J\gtrsim$22--23 mag is dominated by the late-types
(cf. \cite{d95a}, \cite{dwg95}).
There are relatively few early-type galaxies at faint magnitudes. Models
indicate that either luminosity evolution or an extra dwarf population of
late-types is needed to explain the counts of the later types. Redshift
surveys suggest the former, i.e. evolution through episodic starbursts.
The new data and new models presented here do not support the need for 
re-normalizing the total galaxy count models at $b_J\simeq18$ mag. While not
ruling out this need, they show that if this re-normalization is
necessary then it {\it must} be a function of galaxy morphology.
Brighter objects appear larger with the broad trend of increasing apparent size 
as one
goes from early to mid- to late-types. This appears to be true over a range of
almost 10 magnitudes. There is no sharp size cutoff between types. In general,
the early-types are redder, while the later types are bluer with the mid-types
in between. Again, there is no simple way to differentiate types based on
observed colors, especially with {\it only two bands}.

From the current study, we provided the first systematic  $b_J$--band counts 
as a function of galaxy type to address the problem of normalizing the model 
galaxy counts, which use the known local LF's as a function of morphological 
type. 
The galaxy statistics at the bright end ($b_J\lesssim 19$ mag) are still rather
poor. At $b_J\approx 18$ mag, the galaxy counts are approximately 100 galaxies
per 0.5 mag per square degree. A single WFPC2 field is 0.0013 square degrees,
which means that about ten fields are needed to see even one galaxy at random
in this magnitude bin. This implies that a few hundred fields would be needed
to have adequate bright end statistics. Therefore, larger area detectors are
needed. We expect that we can study the magnitude range $16\lesssim b_J 
\lesssim19$ mag from the ground (S. H. Cohen, in preparation), but only in
good seeing, as \eg using
images from the NOAO Deep Wide-Field Survey (\cite{jd99}), and soon also from 
the Millenium Galaxy Catalog (\cite{mgc02}), when classifications are added to
it. This may be feasible because the effective radii at this magnitude are 
larger than the seeing disk in good seeing conditions (see \S~\ref{magre}).
The exact magnitude limit to which these classifications can be reliably pushed 
from the ground is not yet known, but our expectation is to get complete 
and reliable classifications to 
$b_J\lesssim 18-19$ mag. This combined data set will provide better statistics 
at the bright end, which then can be used to more firmly address the 
normalization problem at $b_J=18.0$ mag discussed earlier, and determine if 
the normalization factor is a function of galaxy type.

Some of the requirements to improve upon the interpetation of the faint galaxy 
counts in order to truly get a handle on issues of galaxy formation and
evolution are as follows:
\begin{enumerate}
\item{Brighter galaxy counts ($16\lesssim b_J \lesssim19-20$ mag) as a function of morphological type}
\item{A better handle on the morphological classification accuracy}
\item{Better statistics throughout}
\item{CCD-based type-dependent LF's classified using consistent methods}
\item{Consideration of surface brightness selection effects}
\item{Measured redshifts and better determined k-corrections}
\end{enumerate}

An important point to close on is that we have an incomplete understanding of
the local and intermediate distance Universe. Our knowledge and interpretation
of the distant high-z Universe will always be limited by this. The advent of
bigger and better telescopes brings about the temptation to observe the fainter
and more distant objects in the Universe. These studies still need to be
complemented by those of more nearby objects, such as that which is presented
here, in order to paint the complete picture.

\acknowledgments

This research was funded by NASA grants GO.5985.01.94A, GO.6609.01.95A,
AR.6385.01.95A, \& AR.7534.02.96A (to RAW \& SCO) from STScI, which is operated
by AURA, Inc., under NASA contract NAS5-26555. SHC would like to thank the ASU
NASA Space Grant Graduate Fellowship. We also thank the STScI staff, and in
particular Doug van Orsow, for their dedicated help in getting these parallel
observations scheduled. We also like to thank Drs. J. Bahcall and D. Burstein
for making their respective codes available to us. We also thank the anonymous
referee for their useful suggestions, especially in regards to the completeness
limits. This research has made use of the NASA/IPAC Extragalactic Database 
(NED) which is operated by the Jet Propulsion Laboratory, California Institute
of Technology, under contract with the National Aeronautics and Space
Administration. 

\clearpage

%
%

\clearpage

\clearpage
\begin{deluxetable}{crrrrcrrrrrrr}
\scriptsize
\tablecaption{BBPS Data Summary. \label{tbl-1}}
\tablewidth{0pt}
\tablehead{
\colhead{Field} & \colhead{$\alpha_{2000}$} & \colhead{$\delta_{2000}$} &
\colhead{l$^{(II)}$} & \colhead{b$^{(II)}$} &
\colhead{$A_B$\tablenotemark{2}} & \colhead{F450W\tablenotemark{3,4}} &
\colhead{F814W\tablenotemark{3,4}} & \colhead{SB($\Bj$,lim)\tablenotemark{5}}
& \colhead{$b_{J,lim}$\tablenotemark{6}} & 
\colhead{$I_{lim}$\tablenotemark{5}} \\
\colhead{Name\tablenotemark{1}} & \colhead{($h$ $m$ $s$)} & 
\colhead{($\arcdeg$ $\arcmin$ $\arcsec$)} &
\colhead{(deg)} & \colhead{(deg)} &
\colhead{(mag)} & \colhead{} &
\colhead{} & \colhead{}
& \colhead{(mag)} & \colhead{(mag)}
}
\startdata
bb001\tablenotemark{7} &12:50:11.981 &$+$31:24:53.96 &126.46593 &$+$85.704507 &0.025  &11200(4)	&5100(2)  &24.6  &24.75  &23.75 \nl
bb002  &01:07:13.836 &$+$32:21:15.98 &126.80099 &$-$30.400644 &0.169  &4400(2)	&5600(2)  &24.1  &24.25  &24.25 \nl
bb003  &00:20:13.606 &$+$28:36:18.54 &114.68943 &$-$33.765055 &0.125  &4600(2)	&5600(2)  &24.2  &24.75 &23.75 \nl
bb004\tablenotemark{7}  &12:34:35.321 &$+$07:44:52.51 &290.51752 &$+$70.212661 &0.005  &11200(4)	&5500(2)  &24.1  &24.75  &23.25 \nl
bb005  &23:19:50.369 &$+$08:05:59.64 & 87.49011 &$-$48.367645 &0.153  &5600(2)	&5600(2)  &24.3  &24.75  &24.25 \nl
bb006  &11:17:43.549 &$+$44:18:09.89 &164.38333 &$+$64.541796 &0.000  &4800(2)	&10500(3) &24.0  &24.75  &24.75 \nl
bb007  &23:25:06.940 &$-$12:15:02.39 & 65.03796 &$-$64.892020 &0.073  &11600(4)	&5800(2)  &24.5  &25.25  &24.25 \nl
bb008  &22:56:55.450 &$-$36:35:17.19 &  4.40599 &$-$64.031413 &0.000  &1900(2)	&7500(3)  &23.5  &24.25  &23.75 \nl
bb009  &10:02:24.069 &$+$28:50:05.65 &200.15747 &$+$52.838381 &0.057  &5600(2)	&8400(3)  &24.3  &24.75  &24.75 \nl
bb010  &11:17:29.177 &$+$18:12:37.26 &230.40198 &$+$66.615733 &0.000  &3500(2)	&5400(2)  &23.8  &24.25  &23.75 \nl
bb011  &13:13:30.877 &$-$19:26:31.27 &310.07117 &$+$43.123824 &0.213  &8400(3)	&4100(2)  &24.2  &24.75  &23.25 \nl
bb012  &01:09:56.803 &$-$02:27:02.22 &133.89849 &$-$64.930090 &0.141  &5600(2)	&4700(2)  &24.2  &24.75  &23.75 \nl
bb013  &21:51:04.643 &$+$28:43:48.64 & 81.76111 &$-$19.377477 &0.369  &6000(2)	&4200(2)  &24.3  &24.25  &23.25 \nl
bb014  &01:10:01.312 &$-$02:24:28.94 &133.92383 &$-$64.882760 &0.141  &11200(4)	&5500(2)  &24.4  &24.75  &23.75 \nl
bb015\tablenotemark{7}  &12:36:12.756 &$+$12:35:01.77 &288.41154 &$+$75.024854 &0.125  &7300(3)	&2900(2)  &24.2  &24.75 &23.75  \nl
bb016\tablenotemark{7}  &12:31:16.915 &$+$12:28:06.02 &284.09600 &$+$74.600629 &0.085  &5600(2)	&5600(2)  &24.2  &24.25  &23.75 \nl
bb017  &10:04:52.212 &$+$05:14:59.54 &234.22682 &$+$44.750990 &0.001  &5400(2)	&4600(2)  &24.2  &24.75  &24.25 \nl
bb018\tablenotemark{7}  &12:25:31.456 &$+$12:57:52.96 &278.48442 &$+$74.593150 &0.109  &7500(3)	&2100(2)  &24.3  &24.75  &23.25 \nl
bb019\tablenotemark{7}  &13:21:41.828 &$+$28:53:29.83 & 49.48354 &$+$83.094088 &0.005  &10900(4)	&5400(2)  &24.5  &25.25  &24.25 \nl
bb020  &01:10:00.052 &$-$02:27:30.75 &133.93187 &$-$64.933420 &0.141  &8400(3)	&5000(2)  &24.1  &24.75  &23.25 \nl
bb021  &12:19:35.983 &$+$47:23:10.10 &137.96173 &$+$68.801852 &0.000  &5800(2)	&4500(2)  &24.4  &24.75  &23.75 \nl
bb022  &10:34:54.818 &$+$39:45:57.58 &180.01970 &$+$59.085801 &0.000  &7700(3)	&2800(2)  &24.4  &24.75  &23.75 \nl
bb023  &00:18:27.140 &$+$16:21:16.09 &111.55027 &$-$45.787178 &0.065  &8400(3)	&4700(2)  &24.2  &24.75  &23.25 \nl
bb024\tablenotemark{7}  &12:23:29.779 &$+$15:51:21.40 &271.60958 &$+$76.996802 &0.037  &5400(2)	&5400(2)  &24.3  &24.75  &23.75 \nl
bb025  &21:07:32.081 &$-$05:22:23.81 & 44.61967 &$-$32.581289 &0.241  &2900(2)	&700(1)  &23.8   & \nodata  & \nodata \nl
bb026  &16:36:34.801 &$+$82:34:11.34 &115.71923 &$+$31.066578 &0.293  &3200(2)	&900(1)  &24.0   & \nodata  & \nodata \nl
bb027  &10:24:38.286 &$+$47:04:36.35 &168.23496 &$+$55.070087 &0.000  &5600(2)	&3000(4)  &24.3  &24.75  &23.75 \nl
bb028  &14:17:43.542 &$+$52:23:20.91 & 96.23854 &$+$60.034761 &0.000  &6200(2)	&4100(2)  &24.5  &25.25  &23.75 \nl
bb029\tablenotemark{7}  &12:56:53.243 &$+$22:06:43.94 &317.08991 &$+$84.833774 &0.125  &6000(2)	&4300(2)  &24.3  &24.75  &24.25 \nl
bb030  &00:49:18.900 &$-$27:52:42.31 &334.92501 &$-$89.114023 &0.025  &12200(4)	&5600(2)  &24.6  &24.75  &23.75 \nl
bb031  &00:49:18.664 &$-$27:52:03.10 &335.35253 &$-$89.122747 &0.025  &9900(3)	&5800(2)  &24.5  &24.75  &24.25 \nl
\enddata
\tablenotetext{1}{These are the field names listed in the order the data was
received. The original data can be obtained from the HST Archive
(http://archive.stsci.edu/hst/search.php) by entering the coordinates in the
search form.}
\tablenotetext{2}{Galactic absorption A$_B$ (in mag) from Burstein \& Heiles (1982).}
\tablenotetext{3}{Total integration times in seconds (number of exposures).}
\tablenotetext{4}{There is a total of about 151 Parallel HST orbits in this data set.}
\tablenotetext{5}{SB($\Bj$,lim) is the limiting surface brightness in $\Bj$
$mag$ $arcsec^{-2}$ for a given field averaged over the 3 WF detectors. The PC
limit is approximately 1.5 $mag$ $arcsec^{-2}$ brighter.}
\tablenotetext{6}{Center of faintest complete 0.5-mag bin in the total 
galaxy counts (50\% complete).}
\tablenotetext{7}{Field is in or near the Coma or Virgo superclusters
(see  \S 3.6 and \S 3.7 for details).}
\end{deluxetable}

\clearpage
\begin{deluxetable}{llcccc}
\tablewidth{0pt}
\tablecaption{Differential $\Bj$-band Galaxy Counts 
as a Function of Type \tablenotemark{1} \label{tbl2}}
\tablehead{
\colhead{$\Bj$} & \colhead{$\Bj$} & \colhead{$\log_{10}(n)$} &
\colhead{$\log_{10}(n)$} & \colhead{$\log_{10}(n)$} & 
\colhead{$\log_{10}(n)$} \\
\colhead{lower \tablenotemark{2}} & \colhead{upper \tablenotemark{2}} & 
\colhead{Total \tablenotemark{4}} & \colhead{E/S0}  & \colhead{Sabc}  &
\colhead{Sd/Irr}
}
\startdata
18.5 &19.0  &1.431  &1.431 &\nodata &\nodata \nl
19.0 &19.5  &1.732  &1.431  &1.431 &\nodata \nl
19.5 &20.0  &2.209  &1.732  &1.908  &1.431 \nl
20.0 &20.5  &2.033  &1.431  &1.431  &1.732 \nl
20.5 &21.0  &2.473  &2.033  &1.732  &2.130 \nl
21.0 &21.5  &2.687  &2.033  &2.033  &2.431 \nl
21.5 &22.0  &3.011  &2.334  &2.607  &2.607 \nl
22.0 &22.5  &3.164  &2.431  &2.635  &2.878 \nl
22.5 &23.0  &3.390  &2.577  &2.878  &3.121 \nl
23.0 &23.5  &3.596  &2.753  &3.172  &3.276 \nl
23.5 &24.0  &3.840  &2.863  &3.350  &3.596 \nl
24.0 &24.5  &4.006  &2.857  &3.556  &3.764 \nl
24.5 &25.0  &4.162  &3.021  &3.609  &3.974 \nl
\enddata
\tablenotetext{1}{All counts are $\log_{10}$(Number/sq. degree/0.5 mag)}
\tablenotetext{2}{Lower and upper $b_J$-magnitude bounds of given bin}
\tablenotetext{3}{Area is 0.0370 square degrees, except in last 2 bins, where
the effective areas are 0.0361 sq. deg. and 0.0295 sq. deg., respectively, as
discussed in \S3.7}
\tablenotetext{4}{The $23\lesssim b_J \lesssim 23.5$ mag bin is the faintest
bin where the total counts are better than 90\% complete.}
\end{deluxetable}

\end{document}